# Dead, Slow and Overworked Graphite: Operando X-ray Microdiffraction Mapping of Aged Electrodes

G. Oney[a], F. Monaco[a], S. Mitra[a], A. Medjahed[b], M. Burghammer[b], D. Karpov[b,e], M. Mirolo[b], J. Drnec[b], I.C. Jolivet[c], Q. Arnoux[d], S. Tardif[e], Q. Jacquet[a,*], S. Lyonnard[a,*]

*[a]Univ. Grenoble, Alpes, CEA, CNRS, Grenoble INP, IRIG, SyMMES, F-38000 Grenoble*
*[b]European Synchrotron Radiation Facility, F-38043 Grenoble Cedex, France*
*[c]TotalEnergies OneTech, CSTJF, avenue Larribau, 64018 Pau Cedex, France*
*[d]TotalEnergies OneTech, Centre de recherche de Solaize, 69360 Solaize, France*
*[e]Univ. Grenoble Alpes, CEA, IRIG, MEM, 38000 Grenoble, France*

*[*]Corresponding authors:*

quentin.jacquet@cea.fr
sandrine.lyonnard@cea.fr

## Abstract

Aging limits lithium-ion battery lifetime and must be understood to improve durability and performance, requiring a detailed understanding of how aging alters the availability of cyclable lithium and the integrity of active particles. In this work, (de)lithiation mechanisms are examined and spatially-resolved at the microscale in aged graphite electrodes dismounted from a large format graphite/$LiFePO_4$-$Li(NiCoAl)O_2$ cell at 70% remaining capacity. A multi-technique workflow is employed, combining electrochemical methods with post-mortem structural and morphological analyses, and introducing synchrotron microX-ray 2D diffraction imaging as a technique to probe aged graphite, applied at C-rates from C/5 to C. In-plane and through-plane heterogeneities in graphite dynamics are evidenced, showing the presence of inactive regions localized in two dimensions. In these areas, particles are either disconnected (irreversibly lost) or kinetically limited (reactivated at a slow C-rate), with dead or slow particles exhibiting a wide range of compositions, from x = 0 to x = 1 in $Li_xC_6$. These inactivated graphite particles are found to be heterogeneously distributed throughout the depth of the aged negative electrode. In particular, the most inactivated region localizes at the negative electrode-separator interface, correlating to overworking graphite near the separator.

## 1. Introduction

The demand for improved lithium-ion (Li-ion) battery technologies—specifically, the need to prolong battery life for daily use—drives the necessity to understand and mitigate capacity loss over a battery's life cycle.[1–3]. Graphite (Gr) is the most widely used negative electrode material while at the positive side materials such as layered transition metal oxides ($LiMO_2$; M = Co,

Ni, Mn, Al) or $LiFePO_4$ (LFP) are used. Aging in Li-ion batteries is attributed to two main effects (1) loss of cyclable lithium (LCL) and (2) loss of active material (LAM), which both contribute to increased cell polarization and capacity loss.[4,5] Cyclable lithium loss corresponds to the irreversible trapping of Li inside the electrode particles, in the solid electrolyte interphase (SEI) at the graphite-electrolyte interface, or as inactive Li metal if Li plating occurs. Active material loss occurs when electrode particles no longer participate to the electrochemical reaction. This can be caused by the mechanical disconnection of the particle from the conducting network after repeated volume expansion/contraction cycles due to Li insertion/removal, or due to the formation of thick resistive SEI layers that impede ion and electron transport.

LCL and LAM can be easily distinguished using electrochemical methods such as dVdQ analysis, as demonstrated by Lewerenz et al.[6] Most cells using graphite as a negative electrode reach end-of-life (*i.e.* 80% of initial capacity) due to LCL.[7,8] Loss of active graphite during aging varies between 2 – 30 wt% depending on the cell chemistry and aging protocol,[9–11] but it is rarely the direct capacity-limiting factor because graphite is added in excess during cell fabrication to prevent Li plating.[12] However, aging mechanisms are often intertwined, and inactive graphite can contribute to LCL through several pathways. First, Li can become trapped within inactive graphite particles, as demonstrated by Mikheenkova et al.[13]. Second, the presence of inactive graphite may cause local Li plating due to insufficient Li intercalation sites at the graphite electrode, as suggested by Ko et al.[14,15]. Finally, inactive particles in the electrode can no longer transport Li ions, leading to increased of the electrode tortuosity as Li ions must diffuse around these inactive components.

Therefore, reducing the loss of active graphite is crucial to prevent battery aging. However, understanding the origin of active graphite loss is challenging because it depends on the aging conditions, and the most commonly employed electrochemical methods do not provide information on the localization of inactive particles, their Li concentration, or the presence and thickness of the SEI layer. Moreover, electrochemical measurements are typically performed during low-rate charging check-up cycles, and therefore only quantify active material loss under low-current conditions. However, particles that appear active at low currents may become inactive under higher currents due to highly resistive surface layers or hindered electronic conduction pathways.[16] All these factors contribute to the necessity of using complementary techniques to study the various aspects of active graphite loss in detail.

Scipioni et al.[17] observed what appeared to be electronically disconnected graphite particles at the electrolyte/electrode interface of an aged graphite electrode using ex situ cross-section scanning electron microscopy (SEM). However, they detected these disconnected particles by observing their charging under the electron beam, without confirming the disconnection through complementary techniques. Mikheenkova et al.[13] used ex situ scanning X-ray diffraction to measure the in-plane heterogeneity of a graphite electrode. In the discharged state, they observed some lithiated particles, particularly near the current collector tabs, which could potentially suggest that these particles are inactive. However, such conclusions remain difficult to confirm using static ex situ measurements alone. Waldmann et al.[18] used glow discharge optical emission spectroscopy depth profiling to measure the Li concentration through the depth of aged graphite electrodes. The authors observed that Li appears to accumulate more at the separator surface compared to the current collector side of the electrode during aging, possibly due to the presence of a thick SEI and Li trapping in the graphite particles. However, the authors did not perform complementary diffraction experiments to definitively identify the presence and nature of the lithiated graphite stages. Overall, relatively little is known about graphite inactive particles, presumably due to the inherent limitations of current characterisation techniques.

Recently, operando micro X-ray diffraction (µXRD) has demonstrated its ability to depth-resolve lithiation mechanisms at the micron scale in pristine graphite[19,20], providing insights beyond traditional electrode-averaged and/or post-mortem characterization methods. Several teams have revealed significant spatial heterogeneities during cycling. Tardif et al.[21] quantified lithium heterogeneities across a graphite electrode during delithiation with 1 µm resolution, revealing that reduced (de)intercalation kinetics at high stoichiometries explain the unexpectedly homogeneous behavior observed during the $LiC_6$-$LiC_{12}$ phase transition. Finegan et al.[22] used operando X-ray diffraction depth profiling during exteme fast charging, revealing important state-of-charge (SoC) and lithiation stage heterogeneities across thick graphite electrodes, and quantifying lithium plating localized within first 15-20 µm from the separator. These studies have shown that Li concentration gradients develop through the electrode thickness, with asymmetric (de)lithiation behavior and local current density gradients that depend on both the C-rate and cycling conditions. These operando observations have provided critical experimental validation for porous electrode models and significantly advanced our understanding of Li transport and intercalation dynamics in graphite electrodes. However, in aged graphite electrodes where tortuosity, charge transfer kinetics, and active particle fraction are altered, such spatially resolved studies under operando conditions remain unexplored, despite being equally important for refining aging simulations and models.

In this work, we address the lack of spatially resolved understanding of graphite degradation by investigating aged graphite electrodes extracted from a commercial $LiFePO_4$-$Li(NiCoAl)O_2$//Gr cylindrical cell at 70% of remaining capacity. We employ a multi-modal characterization workflow, combining ex situ electrochemical analysis, X-ray diffraction and nanotomography with operando scanning µX-ray diffraction (µXRD). This operando µXRD approach provides micron-scale, time and depth-resolved structural maps of $Li_xC_6$ phases during cycling, enabling direct observation of lithiation dynamics in both pristine and aged elecrodes.

Our results reveal significant differences in the (de)lithiation mechanisms between aged and pristine graphite electrodes across different cycling rates from slow (C/5) to fast (1C). We provide direct quantification and spatial mapping of inactive graphite in aged electrodes, identifying both inactive lithiated and non-lithiated particles. Non-lithiated inactive graphite particles are found near the separator, while lithiated inactive particles are mainly located in the central region of the electrode. When the behavior of electrochemically active regions is isolated, we find that the characteristic top-to-bottom phase propagation observed in fresh electrodes is no longer seen in aged ones. Multiple graphite phases coexist heterogeneously at the micrometric scale in aged electrodes.

## 2. Results and Discussion

The integrated workflow developed to systematically characterize aged and pristine graphite electrodes is shown in Figure 1. The cylindrical element built with double-side coated graphite and $LiFePO_4$ (LFP)/$Li(NiCoAl)O_2$ (NCA) electrodes (90/10 w%), was aged by cycling at a C/2 rate (one full charge or discharge every two hours) at room temperature, within a voltage range of 2.5–3.8 V. Under this voltage window, the electrochemical activity is largely due to the LFP active material in the blend positive electrode and the exceptional cycling stability of LFP[23,24] allows us to isolate the information on graphite aging. Over approximately 3000 cycles, the element lost 30% of its initial capacity (determined by a control cycle at C/3 rate), indicating it had reached a state beyond the EoL defined using the 80% capacity criterium. This state-of-health (SoH) reflects a realistic and industrially relevan t condition, making it well-suited for studying degradation mechanisms under a defined aging scenario.

Afterward, the cell was fully discharged (state of charge (SOC) = 0%) and dismantled in an argon-filled glovebox. The unrolled electrodes were washed three times with dimethyl carbonate (DMC) and dried. Rectangular sections from the center of the unrolled aged electrodes were cut (Figure 1a). One side of the double coating was removed in the glovebox, mechanically for the graphite electrode, and removed with NMP for the positive electrode. Electrode disks of varying sizes were punched from these aged and pristine (i.e., uncycled) electrodes based on the requirements of each characterization technique (see holes in the photo of Figure 1a, top), e.g., disks for coin cells ($\phi$ = 15 mm) or miniature Swagelok-type cells ($\phi$ = 1 and 3 mm, used in synchrotron experiments) (Figure 1b). For the electrochemical characterizations, the pristine electrodes underwent defined formation cycles in the coin cell to create a stable solid-electrode interphase layer (see Experimental Section). For simplicity, we refer to them as “pristine” electrodes throughout the manuscript, although all characterizations were performed on “pristine formed” electrodes. Over 30 coin cells were cycled using pristine and aged electrodes to evaluate the electrochemical performance decay, ensuring representative comparison with operando cells. Multiple post mortem and operando experiments were conducted using electrochemical and X-rays-based techniques in the lab and at the synchrotron (Figure 1c). Data were correlatively analyzed to couple electrochemical, chemical, structural, and morphological information gained at several scales at varying SoCs at equilibrium and out-of-equilibrium states (Figure 1d). Details on materials preparation and experimental set-ups are provided in the Experimental Section, and a discussion of sample integrity during electrode handling is included in the Supplementary Information.

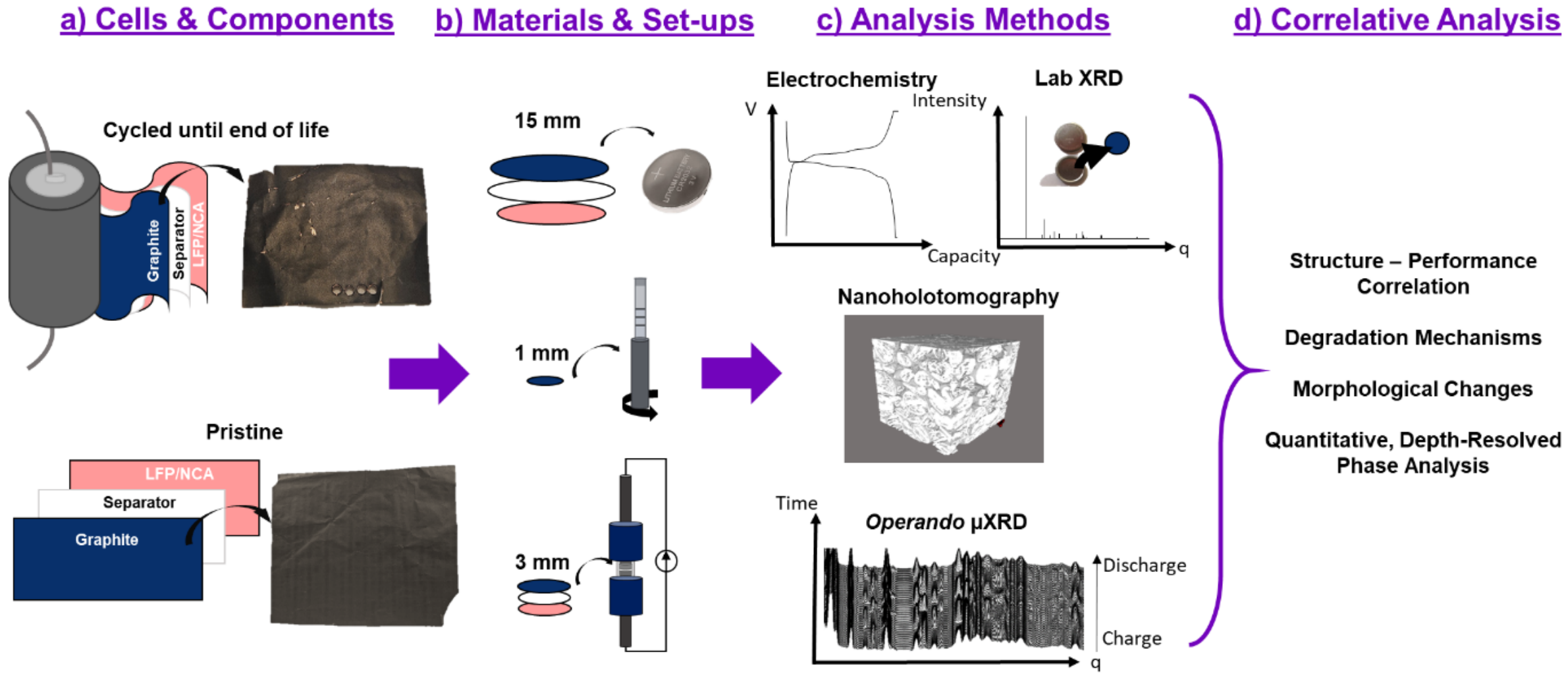

**Figure 1.** Integrated workflow to characterize aged and pristine graphite electrodes. **(**a) Schematic representation of the cylindrical cell and the pristine electrodes used in this study. Photographs show a cycled graphite electrode with post mortem punching holes (top) and a pristine graphite electrode (bottom). (b) Electrode disks of varying sizes punched from aged and pristine electrodes and mounted in coin cells or miniature capillary-type cells based on the requirements of each characterization technique. (c) Ex situ and operando analysis methods. Top: electrochemical data, and post mortem X-ray diffractogram on directly punched or re-cycled electrodes extracted from coin cells. Middle: 3D reconstruction of graphite electrodes showing distinct phases (particles in white and pore-binder-SEI in grey). Bottom: typical time-resolved XRD patterns obtained by scanning operando µXRD in Swagelok-type cells. (d) List of correlative data analysis used in this work to identify the degradation mechanisms.

### 2.1. Ex Situ Analysis of Electrochemical Performance and Morphology Evolution

To evaluate the materials' capacity and galvanostatic behavior, electrode disks were punched from pristine and aged electrodes and reassembled in coin cells against pristine and aged LFP/NCA (Figure 1a-c). These cells underwent a controlled (dis)charging sequence, followed by electrochemical analysis to examine key reaction steps (see Experimental Section). At a C/5 cycling rate between 2.5–3.8 V, the discharge capacities at room temperature were $Q_{disc,aged}$ = 1.66 ± 0.32 mAh/cm² in average for 9 aged full cells and $Q_{disc,pristine}$ = 2.26 ± 0.14 mAh/cm² for 25 pristine cells tested (Figure 2a) confirming the degradation in electrochemical performance observed in the cylindrical element ($Q_{disc,aged} = Q_{disc,pristine}$ x 73%).

Further insights are gained using differential voltage analysis (DVA, i.e., dV/dQ plots), an electrochemical method that allows the discrimination between lithium inventory and active material loss.[25,26] In DVA, peaks correspond to inflection points of the electrochemical potential during (de)lithiation. The graphite (de)lithiation mechanism follows a well-documented staging process (stages n = 1 to 4, *n* represents the number of graphene sheets between two interlayer spaces containing Li-ions). This process produces a characteristic electrochemical potential curve with three plateaus corresponding to solid solution mechanisms separated by the various phase transition regions ($Li_xC_6$ phases, *x* being the lithium content). The stoichiometry of stages 2 and 1 ($LiC_{12}$ and $LiC_6$, respectively) is known, while not precisely defined for stages 2L, 3L and 4L due to liquid-like in-plane ordering.[27] LFP undergoes a biphasic reaction during (de)lithiation, producing a flat electrochemical profile [28], while in NCA solid-solution reactions govern the process[29,30]. Note that, within our full cell cut-off voltage (3.8 V), the primary contributing positive electrode is LFP. The DVA of a full cell typically contains peaks originating from the negative and the positive electrodes. Indeed, the pristine full cell shows five peaks at 0.4, 0.55, 1.45, 2, 2.25 mAh.cm² (Figure 2b, top panel), corresponding to phase transitions in our system.

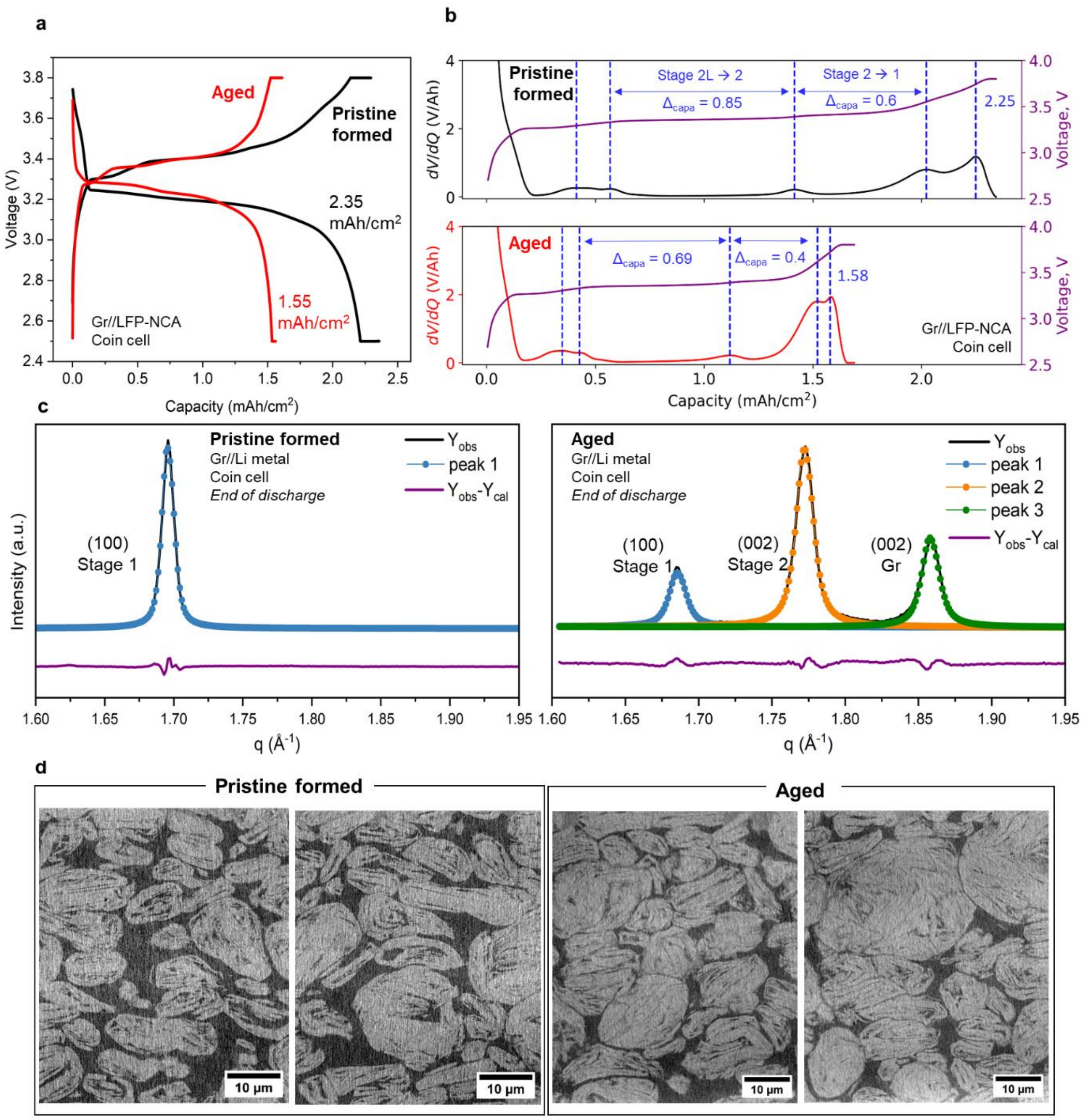

**Figure 2.** Ex situ characterization of graphite. (a) Galvanostatic charge-discharge profiles at a C/5 rate for pristine and aged graphite reassembled in coin cells paired with pristine or aged LFP/NCA electrodes extracted from the same cylindrical cell. (b) Derivative voltage analysis of pristine and aged full cells measured at a C/5 rate. (c) graphite electrodes discharged to 0.005 V for graphite lithiation in a half cell configuration at a C/20 cycling rate. (d) 2D images of the graphite electrode measured by synchrotron nanoholotomography, comparing pristine (after

formation cycle in coin cell) and aged (extracted from the element at discharged state) samples. The images have a pixel size of 25 nm and a field of view of 50 µm. For each sample, two different slices are shown. Separator side is the upper part of the shown slices. Dark regions correspond to the pore-binder-SEI phase, while lighter regions represent graphite particles.

To disentangle the contribution from each electrode, DVA analysis of pristine graphite/Li and LFP-NCA/Li half cells was conducted (Figure S1 and S2, Supporting Information). DVA of graphite/Li shows three main peaks, a large one composed of several peaks centered on 0.57 mAh.cm² and two narrow peaks at 1.68, 3.2 mAh.cm². The large peak corresponds to the successive formation of stage 4, stage 3, and stage 2L, having estimated composition of $LiC_{36}$, $LiC_{30}$, and $LiC_{18}$, while the narrow peaks are attributed to $LiC_{12}$ (stage 2) and $LiC_6$ (stage 1), respectively. DVA analysis of LFP-NCA shows three peaks at 2.25, 2.40, and 2.50 mAh.cm² mostly due to NCA since the LFP electrochemical profile is flat.

Based on these results, the DVA of the pristine full cell (Figure 2b, top panel) can be readily interpreted. The three first peaks at 0.40, 0.55, 1.45 mAh.cm² correspond to the formation of stages 3, 2L, and 2, while the last ones at 2.00 and 2.45 mAh.cm² are attributed to the NCA. Only two peaks are observed for the NCA rather than three compared to half cells due to lower cut-off voltage. Note that stage 1 formation is not observed in full cell due to the electrode balancing (i.e., the electrodes are never fully lithiated), while it is detected in half cell configuration (Figure 2c). As seen in Figure 2b-bottom panel, DVA of the aged cells features the same five peaks but at different capacity values of 0.30, 0.45, 1.10, 1.51, 1.58 mAh.cm². There is a shrinkage in the plateau between two NCA peaks indicating active material loss in the positive electrode. The apparent shrinkage in capacity between stages 2L and 2 indicates a loss of active graphite, estimated to be approximately 20 %. The observation of active graphite loss is consistent, albeit different in value, with the differential voltage and incremental capacity analyses conducted on the graphite/Li half cell (Figure S1, Supporting Information). However, on the full cell analysis, the reduction in delivered capacity on the final plateau is larger than the estimated loss of active graphite. This suggests that while lithium intercalation sites remain available in the aged graphite electrode, lithium is no longer available to intercalate. Thus, while graphite loss is evident in the negative electrode, it is not the limiting factor for capacity in this full cell. The true limiting factor here is the increase in positive electrode voltage during NCA delithiation, as indicated by the presence of two NCA peaks at the end of charge in the DVA analysis. This shift in the NCA peaks is attributed to negative electrode slippage[31]—a change in the portion of positive electrode potential utilized. In fact, the aged positive electrode cycles between 40% and 100% SoC. Such slippage is typically caused by irreversible reactions at the negative electrode, most likely due to the continuous formation of the solid electrolyte interphase (SEI). Additionally, the incremental capacity analysis (ICA) of the full cell (Figure S3, Supporting Information) shows polarization effect, hence a resistance increase in the aged graphite, which is a sign of thick resistive SEI that can impede ion and electron transport and cause both loss of active material and cyclable lithium. Overall, electrochemical methods reveal that both aging mechanisms—loss of active graphite and loss of lithium inventory—are present in this aged graphite.

X-ray diffraction is used to investigate the nature of the graphite active material loss and to confirm Li inventory loss. Details on pattern fitting and peak analysis method are provided in the Experimental Section and Supporting Information (Figure S4, and S5, Supporting Information). First, X-ray diffraction of aged LFP electrodes extracted from the element in the discharged state shows the presence of 40% of the delithiated FP phase, confirming the drift of the positive electrode working potential range due to the Li inventory loss (Figure S4, Supporting Information). On the graphite side, we examined a larger electrode piece (~3 × 4 $mm^2$) directly extracted from the discharged element by XRD. The analysis based on peak fitting revealed a phase distribution of 20.8% lithiated stage 2L/L and 80.2% non-lithiated graphite (Figure S5, Supporting Information) (average $x_{aged,disch,element} \approx 0.1$). This suggests that *some lithiated particles stopped responding to the applied current*.

To further investigate the participation of the active material to electrochemical operation, a pristine electrode and an aged electrode extracted from the cycling element were re-lithiated in half cell configuration at a slow C/20 rate (Figure S6, Supporting Information), followed by a float at 0.005 V with a C/100 current limitation to reach thermodynamic equilibrium. The experiment on the pristine graphite gives 100% of stage 1, as expected (Figure 2c). For the aged electrode, peak profile refinement of the ex situ XRD pattern of the electrode shows the presence of three different phases: namely fully lithiated stage 1 (26.7%), partially lithiated stage 2 (41.0%), and non-lithiated graphite (32.3%) (Figure 2c). Therefore, *some graphite particles in the aged sample are no longer participating to the electrochemical reaction*, and some regions remain in an intermediate state of lithiation, not able to reach $LiC_6$ anymore. This situation can result from the disconnection of graphite particles either in their fully delithiated state, either in poorly lithiated states, and be the sign of irreversible morphological constraints as well as lithium trapping in highly degraded areas due to pore clogging and SEI thickening. But it is also possible that the situation is even more complex, as there is no indication at this stage of the nature and distribution of local lithiation states. Overall, these results indicate that the composition of the aged graphite is complex, with possibly several types of inactivation or inactivity induced by the cell usage until its end-of-use.

A visual comparison of pristine and aged graphite electrodes at delithiated state shows no significant differences (Figure 1a). The aged electrode remains attached to its Cu current collector, with no visible signs of undesired lithiated graphite phases ($LiC_{18}$, $LiC_{12}$ and $LiC_6$ are blue, red and golden, respectively)[32]. Despite the substantial electrochemical inactivity observed, the electrode does not exhibit noticeable physical degradation and retains its overall macroscopic structure.

To evaluate the morphology at a microscopic scale, the 3D microstructure of pristine and aged graphite electrodes was compared using ex situ nanoscale holotomography[33] available at the ID16A beamline of ESRF (Figure 2d). Sample preparation, phase-contrast image acquisition method, as well as segmentation and analysis, are described in the Experimental Section. The aged graphite particles appear larger and more agglomerated than pristine ones. To quantify this observation, the porosity was calculated using a 20×20×20 $\mu m^3$ representative volume. In the pristine electrode, the measured porosity closely matched the initial electrode preparation porosity of 35%, validating the methodology. Only 2% of the porosity is micro porosity – not in directly contact with the electrolyte. Note that the voxel size of 25 nm, i.e. a resolution around 75 nm, doesn't allow to visualize the SEI or the small pores. Under these resolution conditions, in aged electrodes, the overall porosity was found to decrease significantly to an average of 25%. This reduction likely results from repeated charge/discharge cycles causing expansion/contraction if the active materials, leading to particle growth, agglomeration, and electrolyte degradation products deposition on graphite surfaces.

Consistent with the literature, the ensemble of our ex situ observations confirms that graphite aging involves both active material loss and Li inventory loss, revealed by changes at macroscale (capacity loss, polarization, loss of available lithium content) and at the microscale (decreased porosity). We further evidenced the presence of a significant amount of "inactive" graphite particles. However, the interplay of the observed degradation mechanisms during electrochemical operation remains unclear at this stage and deserves additional investigations. Several key questions need to be addressed: (1) which graphite phases still actively lithiate in the aged electrode, and at which extent? (2) Where are the inactive phases located within the electrode? (3) What is the nature of this "inactivity" and is it the origin or the consequence of aging?

### 2.2. The principle of X-ray µdiffraction imaging to localize graphite aging

To quantify degradation heterogeneities, and, particularly, to locate inactive graphite and understand its characteristics, we designed an experiment capable of mapping graphite lithiation in the depth of the aged electrode in real time during charge and discharge at different C-rates. The operando experiment consists in scanning the entire aged graphite electrode placed into a 3 mm Swagelok-type operando full cell assembled with an aged LFP/NCA electrode, using a 3 x 3 µm$^2$ X-ray synchrotron beam (ID13 – ESRF) and collecting the transmitted diffracted X-rays on a 2D detector (Figure 3a). The information on the setup and acquisition details can be found in Experimental Section. A pristine graphite was also measured using the same set-up in a full cell assembled with pristine positive electrode as a comparison between pristine and aged electrodes. Both pristine and aged cells were cycled under the same conditions: one slow cycle at C/5 (C/5 means that the current is applied so that the charge lasts 5h ; I = 33.8 µA), one at C/2 (charge in 2h ; I = 84.5 µA), followed by a C/2 charge and a C fast discharge (0.169 mA). The electrodes were also cycled in coin cells to verify any bias from the operando synchrotron experiment. The representativeness of the electrochemical performance of the Swagelok *vs* coin cells is excellent for the pristine sample (Figure 3b), the discharge capacity at C/5 being 2.14 mAh/cm$^2$ corresponding to a lithiation up to x = 0.83 for $Li_xC_6$ (see Supporting Section 3 for calculation). The aged electrode delivers 1.28 mAh/cm$^{2,}$ meaning a maximum lithiation to x = 0.54 (Figure 3b). The shape of the curve is the same, but there is a noticeable difference between Swagelok and coin cell capacity. This discrepancy is attributed to heterogeneity within the aged electrode, as reflected by the larger error bars in the coin cells from aged full cells.

The electrode is scanned with 3 µm and 100 µm in vertical (z-axis) and horizontal (y-axis) resolution, respectively, giving (y,z) diffraction maps of 35 x 28 pixels for the pristine electrode (Figure 3c, upper panel) and 35 x 31 pixels for the aged electrode. Each diffraction map is acquired in approx. 2 min. The detector image collected at each (y,z) position in the electrode is azimuthally integrated[34] to get a classical powder diffraction pattern at the corresponding (y,z) pixel (Figure 3c, bottom panel). The diffraction maps were masked in order to consider only regions with important amount of active material based on the intensity of the graphite detected (Figure S7, Supporting Information).

The (de)lithiation of graphite was tracked by analyzing the XRD patterns between 1.7 and 1.9 $Å^{-1}$, corresponding to the Q-range where unlithiated and lithiated graphite peaks are found. The initial graphite phase appears at high *q*-values (1.87 $Å^{-1}$, corresponding to the graphite (002) crystal plane), while the peak corresponding to fully lithiated stage 1 is at 1.69 $Å^{-1}$ ($LiC_6$ (001)) (Figure 3d), with intermediate stage 2-2L ($LiC_{12}$-$LiC_{18}$) between 1.77 and 1.80 $Å^{-1}$ and stage 3 ($LiC_{30}$) at 1.82 Å. The peaks were fitted using Prisma software[35] with a pseudo-Voigt profile to extract the peak positions and intensities during cycling (Figure S8, Supporting Information).

Two cells (noted cell 1 and 2) with distinct pieces of the same aged graphite material were measured to ensure reproducibility and assess result representativeness. This approach helps exclude intrinsic deviations from manufacturing or assembly defects and evaluates variability from sampling different locations of the large-format aged cell. As both pieces were taken from the same rectangular section of the large cell, potential effects of aging variations across different parts of the element are minimized. Cell 1 and Cell 2 parameters are reported in Table S1 for a thorough comparison. Cell 1 data are shown in details through the paper (Figures 3 to 8) while Cell 2 experimental data are reported in Supporting Information (Figures S9, S14, S16 and S17).

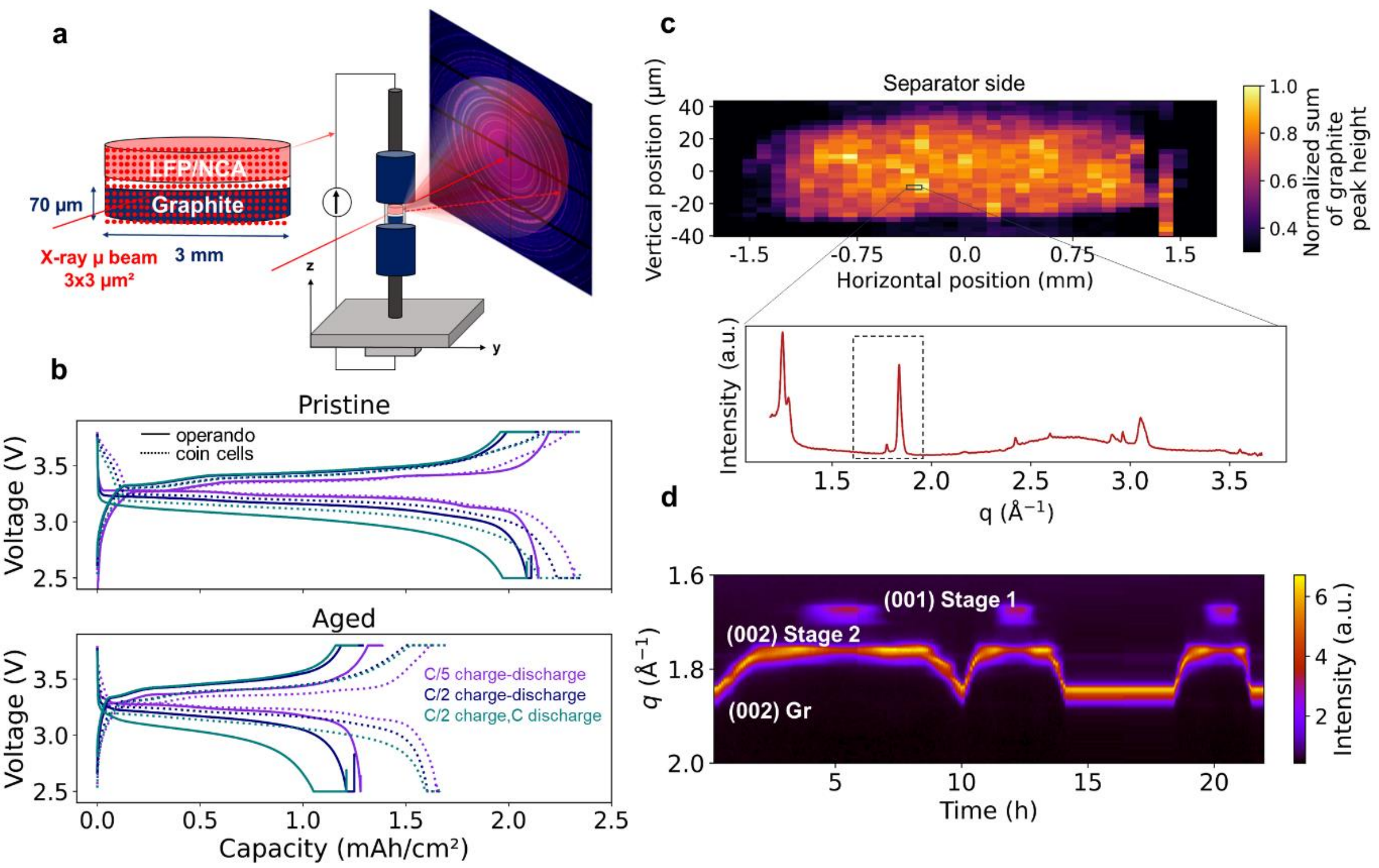

**Figure 3.** Principle of fast operando scanning µdiffraction experiment. (a) Schematic of the operando µXRD setup for probing a full-cell. (b) Galvanostatic cycling curves of pristine and aged graphite//LFP-NCA full cells at various charging and discharging rates. The dots show the electrochemical performance of the materials in coin cells, while the continuous lines show

the performance of the materials in operando cells. (c) Spatial distribution of graphite within the mapped area, based on normalized diffraction peak intensities (1.7 and 1.9 $Å^{-1}$). The inset shows an example of XRD pattern from the highlighted pixel. (d) XRD pattern evolution on the selected pixel over time during electrochemical cycling.

### 2.3. Quantifying and localizing the loss of active graphite particles

We first examine the spatiotemporal evolution of $Li_xC_6$ phase fractions in both pristine and aged electrodes during a C/5 charge-discharge cycle. Phase fractions were extracted from specific areas (averaging 3 x 3 pixels) chosen to explore spatial heterogeneity: Top Left (1), Top Right (2), Bottom Left (3), and Bottom Right (4) (Figure 4a).

As expected, in the pristine electrode, phase evolution is sequential and homogenous across all examined regions (Figure 4b and S10, corresponding diffraction patterns in Figure S12). Stage 3 forms first ($LiC_{30}$), followed by stage 2L/2 ($LiC_{18}/LiC_{12}$) and finally stage 1 ($LiC_6$). At any given time, one phase dominates, and reaches near-complete fraction ~ 100%. Full lithiation to stage 1 is not achieved, which is due to electrode balancing, in line with the lithiation index of x = 0.83 observed from electrochemistry. This process is reversed during discharge, with one dominant phase present at a given time. Such uniform reaction across pixel areas confirms the full participation of graphite particles to the (de)lithiation process.

In contrast, the aged electrode displays distinct regional heterogeneities (Figure 4c). In region 1 (Top Left area in Figure 4a), severe inactivity is observed. Over 60% of the graphite remains completely unlithiated initially. Stage 3 only forms briefly (~10% at t = 0.4h), then gradually transitions to Stage 2/2L and Stage 1. Overall, ~75% of the graphite remains inactive throughout. In this region, there is also a delayed lithiation of ~9% of graphite late in the charge (t = 2-3h), and strong in-plane (along x-axis) heterogeneity. In contrast, the behavior in region 4 (Bottom right in Figure 4a) closely resembles that of pristine electrode. Most of the expected graphite phase transitions take place, with some of the aging mechanisms identified. Residual amounts of unlithiated graphite (9%) and stage 3 (3%) remains at full charge and a delayed behaviour persists as graphite fraction continuously decreases from 15% to 9%. An intermediate degradation behavior is seen in Regions 2 and 3 (Top Right and Bottom Left in Figure 4a, respectively). These zones show coexisting phases since the beginning of the charge and retain more than 25 % unlithiated graphite across the cycle.

Overall, the operando determination of phase fraction evolution in the aged electrode reveals several key features: (1) inactive particles can be stuck in different states of charge with x ranging from 0 to 1 in $Li_xC_6$, and (2) there are also delayed particles lagging behind ensemble electrochemistry. Moreover, a substantial in-plane (x-axis) and out-of-plane heterogeneity (yz-plane) is observed.

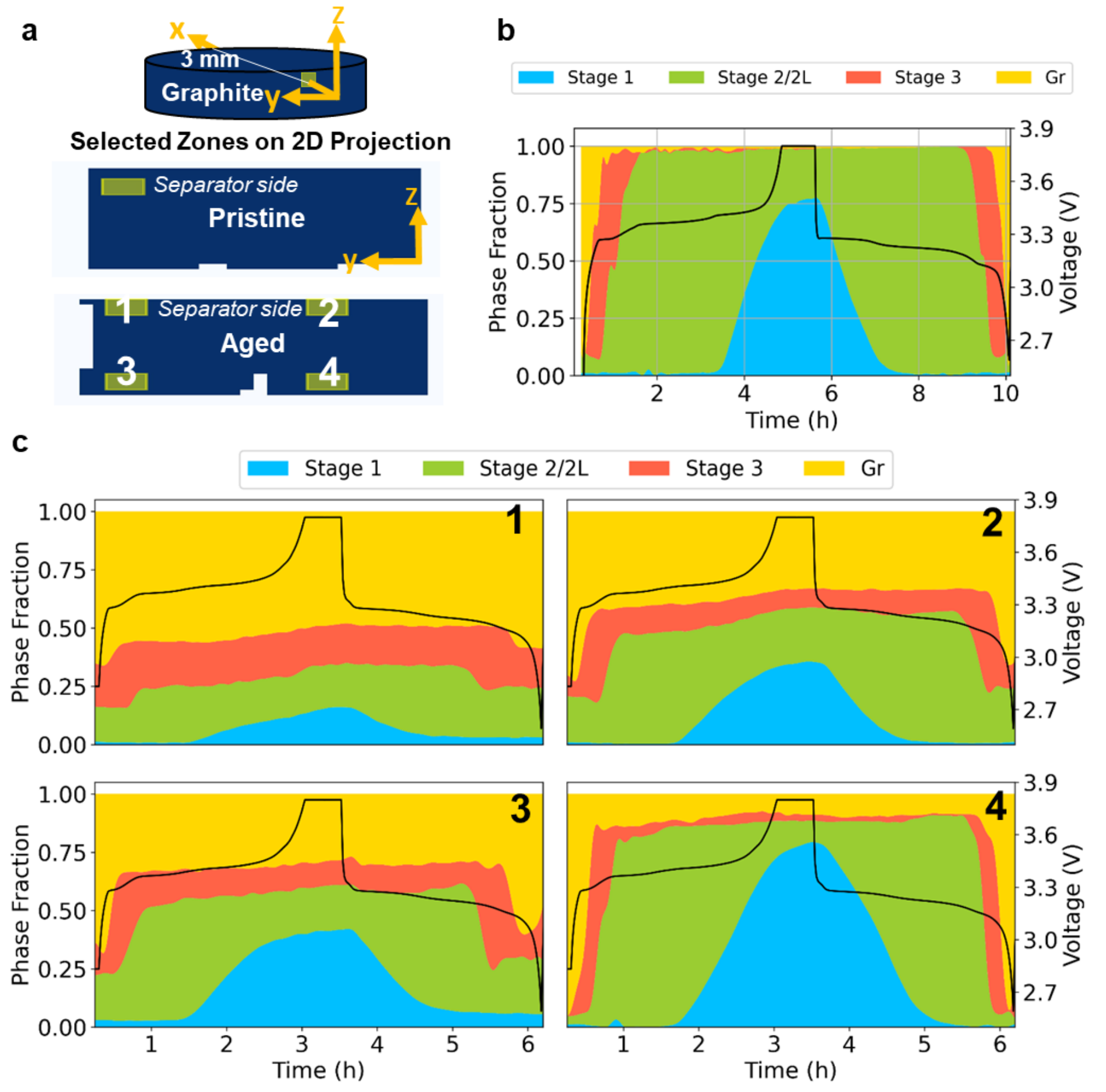

**Figure 4.** Spatiotemporal distribution of $Li_xC_6$ phase fractions in pristine and aged graphite. (a) Schematic of the analyzed area on the graphite electrode. Phase evolution is averaged along the x-axis (direction of the X-ray beam) over representative 3×3 pixel regions (highlighted in green in the 2D projections). Four regions are selected for aged electrode: top left (1) and top right ( near the separator) (2), bottom left (3), and bottom right (4) (near the separator). (b) Phase fraction evolution during C/5 charge and discharge in the selected pristine electrode region. Each color represents a graphite staging: non-lithiated graphite is shown in yellow, stage 3 in coral, stage 2/2L in green, and stage 1 in blue. (c) Phase fraction evolution during C/5 charge and discharge in the four selected zones of aged graphite. The color scheme is consistent with panel (b) for direct comparison.

There is a significant proportion of electrochemically "inactive" graphite in the aged electrode. To further quantify this inactivity and investigate how it spatially distributes in the depth of the electrode, we isolated the "inactive" and "active" graphite populations and generated spatially- and time-resolved maps. The amount of inactive phase is extracted from the X-ray patterns by selecting the peaks that do not change with time during a charge/discharge cycle (or all three charge/discharge cycles), as shown in Figure 5a, where the unchanged peak regions are highlighted in colors. Mathematically, this is performed by applying a minimal filter to the 4D matrix of the diffracted intensities (q, y, z, and time) producing a 3D matrix of the inactive phase patterns (q, y, z). Details on this analysis can be found in Supporting Information (section 4). This is done in each pixel (Figure S13). For each pixel, the concentration and nature of the various inactive phases is determined by fitting the inactive phase pattern to identify the graphite staging regions (1, 2/2L, 3 and 4), and extract the fraction of its population, as done on the raw patterns. A range of lithiated and unlithiated phases was detected, indicating that inactivity is chemically complex (Figure 5a-b). Global percentages of inactive phases and their variations with C-rate are reported in Table 1.

**Table 1.** Calculated percentages of inactive phases with respect to the total amount of graphite. Values are calculated for the entire operando cycling protocol (all cycles) and for each individual C-rate of aged graphite cells 1 and 2. The total inactivity sums the populations of lithiated (stage 1, 2/2L, stage 3) and unlithiated (Gr) particles.

| | | **Total inactivity (%)** | **Stage 1 (%)** | **Stage 2/2L (%)** | **Stage 3 (%)** | **Lithiated inactive total (%)** | **Gr (%)** |
|---|---|---|---|---|---|---|---|
| **Aged Cell 1** | **All cycles** | **34** | **0.7** | **12.4** | **4.4** | **17.5** | **16.6** |
| | **C/5** | 37 | 1.0 | 12.5 | 5.0 | 18.5 | 18.7 |
| | **C/2** | 44 | 1.6 | 17.5 | 7.2 | 26.3 | 17. 9 |
| | **C/2 - C** | 47 | 3.0 | 18.1 | 7.1 | 28.2 | 18.7 |
| **Aged Cell 2** | **All cycles** | **45** | **2.5** | **11.2** | **15.2** | **28.9** | **15.7** |
| | **C/5** | 47 | 2.5 | 11.3 | 15.3 | 29.1 | 17.6 |
| | **C/2** | 54 | 5.2 | 15.5 | 16.9 | 37.6 | 16.3 |
| | **C/2 - C** | 57 | 7.6 | 16.7 | 16.8 | 41.1 | 15.5 |

In the pristine state, inactive regions are absent (less than <1%). In the aged graphite electrode the global averaged inactivity is significantly higher, reaching 34% in cell 1 (45% in cell 2) when cumulating data from all cycles. Typically, in cell 1 (2) 16.6% (15.7) inactive graphite is pure graphite while 17.5% (28.9) is in a lithiated state, most of it being stage 2/2L (stage 3).

We observed that both the amount and nature of inactive phases depend on the C-rate. Higher cycling rates lead to an increase in the overall fraction of inactive material, and a relative increase in lithiated phases within the inactive regions. Interestingly, the faster cycling rates do not generate more inactive unlithiated graphite. This signifies that there is some dead unlithiated graphite that is not recoverable by the slower cycling rates, and some graphite particles are delayed in their lithium intake/outtake with C-rate increase. These findings are consistent with the initial ex situ coin cell analysis, which showed higher capacity when cycling at a slower C/20 rate (Figure S6, Supporting Information). However, while it is clear that some slow graphite can be reactivated at lower currents, it remains uncertain whether all inactive regions can be fully recovered with slower cycling.

Similar trends are observed in the second aged operando cell (cell 2), which exhibited even higher levels of inactivity across all C-rates, suggesting that spatial heterogeneity within the original electrode sheet may play a role in degradation severity.

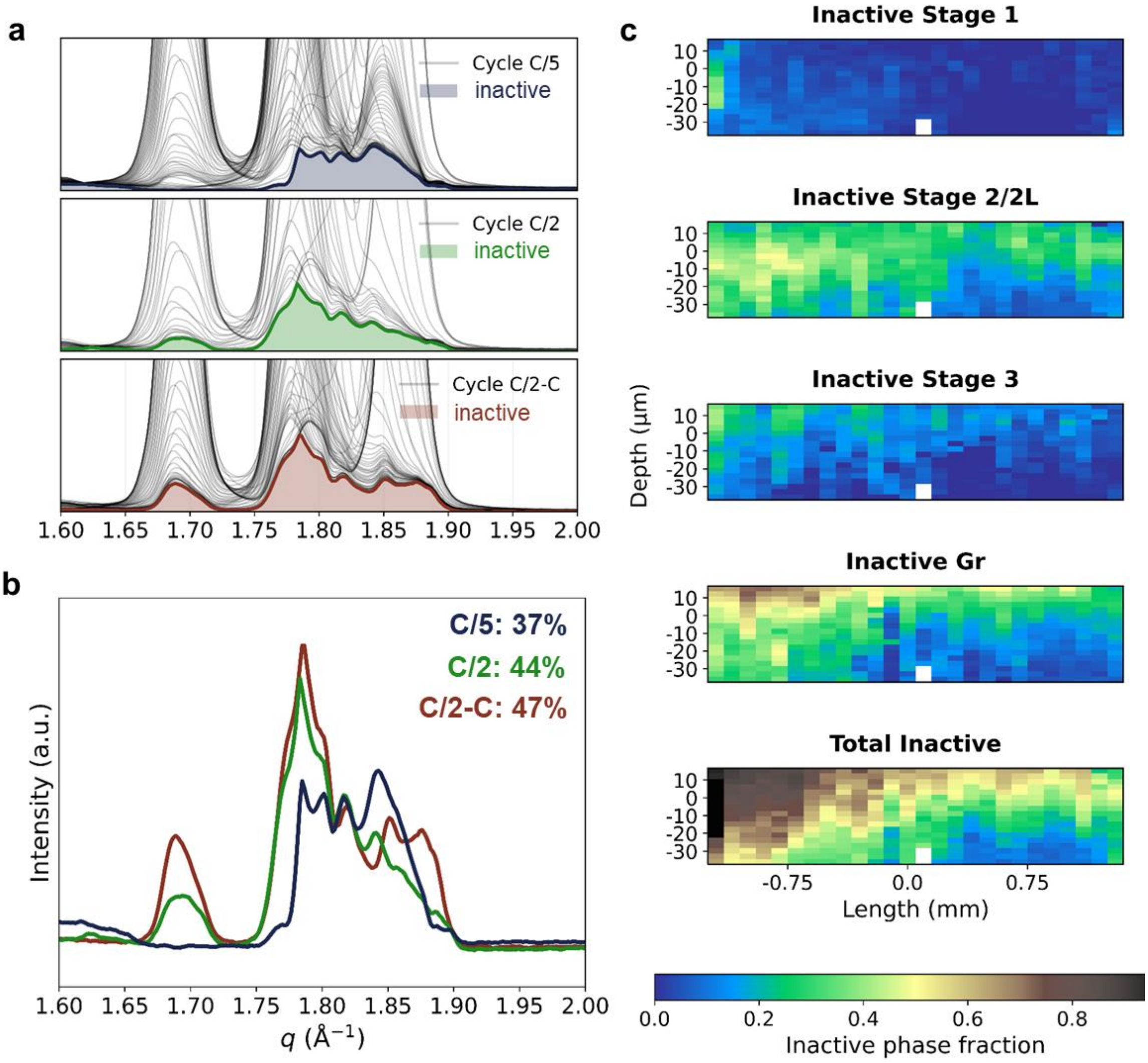

**Figure 5.** Spatial distribution and quantification of inactive $Li_xC_6$ phases in aged graphite electrode. (a) Time-resolved $Li_xC_6$ diffraction peaks evolution during cycling at C/5 (top), C/2 (middle), C/2-C (bottom), with unchanged fraction corresponding to inactive graphite in blue, green and red, respectively. Data are averaged in (y,z). (b) Spatio-temporally averaged XRD patterns of aged graphite under different cycling protocols, showing increased phase inactivity: 37% in C/5 (blue), 44% in C/2 (green), and 47% in C/2 charge-C discharge (red). (c) 2D inactivity phase maps averaged over the full operando cycle. In each pixel, the fraction of inactivity is represented using the color code (Blue : totally active ; dark brown= fully inactive), for the various types of inactive phases : Stage 1 (top), 2/2L and 3 (middle top), pure graphite (middle bottom), and averaging all phases (bottom).

Regarding the spatial distribution of inactive phases, they are generally more concentrated at the top of the electrode (facing the separator) (Figure 5c). Additionally, the left side of the electrode exhibits more inactivity than the right. However, this left/right heterogeneity is less pronounced in the second operando cell (cell 2), while the top/bottom difference remains consistent and reproducible (Figure S16, Supporting Information). This indicates that left/right heterogeneity is a specimen-dependent feature, connected to large scale sample heterogeneities within the electrode itself, while the depth effect is mechanistic. A closer examination of individual phase distributions reveals that unlithiated graphite is primarily located at the outer surface of the electrode, whereas the highest concentration of inactive stage 2/2L is found toward the center. Moreover, the distribution of inactive phases does not depend on C-rates (Figure S15), indicating that cycling rate does not significantly affect their localization.

In conclusion, the activity of aged graphite particles is highly dependent on their location within the electrode, with regions closer to the separator exhibiting greater inactivity. Importantly, some particles trapped in different stages also became unresponsive with aging. This behavior is kinetically dependent, as slow cycling rates improve transport properties and enable some graphite to be active in electrochemistry again. However, the complete reactivation of all aged particles remains uncertain as slower cycling rates on ex situ analysis on half cells did not enable full lithiation to stage 1 phase. This proves that some regions are probably irreversibly disconnected and inaccessible to reactions (dead aged graphite area), while others are hard to reach for ions/electrons and only kinetically-limited (slow aged graphite area).

### 2.4. Correlating graphite inactivity and Li concentration

In light of the previous observation showing large inactive phase fraction heterogeneities, the impact on Li concentration in both the active and inactive phase was investigated. Direct determination of lithium content in graphite using Rietveld refinement was not possible due to the low scattering cross section of X-rays with lithium. Instead, we quantified Li concentration indirectly by analyzing the intensity and position of diffraction peaks in the 1.7–1.9 $Å^{-1}$ range, following the method used in our previous work based on lattice parameter-lithium concentration relationship.[21]

Comparing the indirect Li concentration quantification for the entire electrode with the expected Li concentration obtained from electrochemistry during cycling at C/5 (Figure 6a), we observe a good agreement between $0 < x < 0.3$ and $0.6 < x < 0.9$ for x in $Li_xC_6$. The $0.3 < x < 0.6$ region is not well described because stage 2 and 2L phases corresponding to $LiC_{18}$ and $LiC_{12}$ share the same peak position at 1.78 $Å^{-1}$. For the pristine electrode, XRD data shows that the maximum lithiation corresponds to a global (electrode-averaged) $x = 0.809$ for $Li_xC_6$, while for the aged electrode, this value decreases to 0.654. Both extracted values are consistent with the electrochemical results, with $x = 0.83$ and $x = 0.59$ at the end of the first charge at a C/5 cycling rate, validating the method.

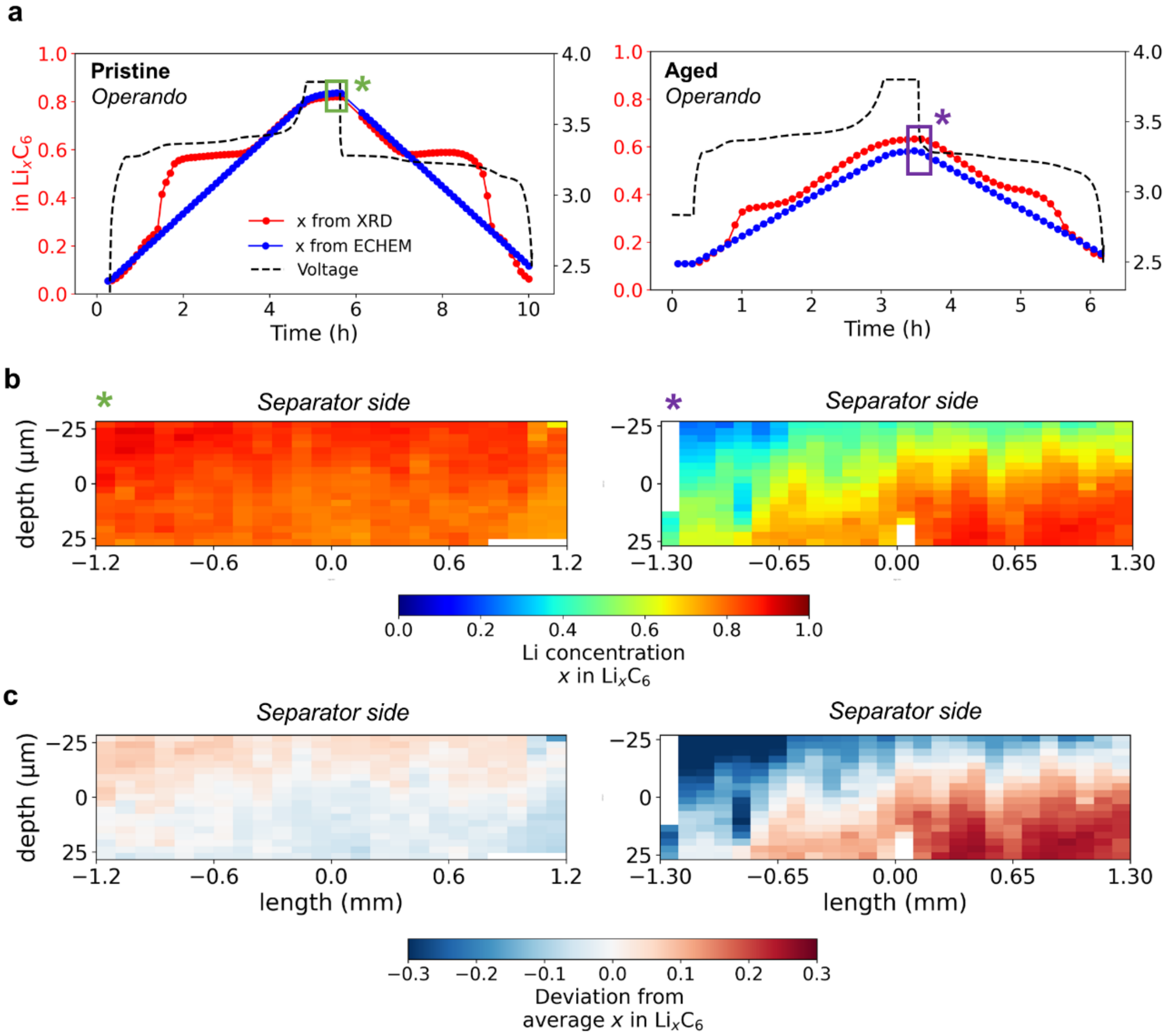

**Figure 6.** Lithium concentration heterogeneity in the depth at end-of-charge for pristine and aged graphite. (a) Evolution of overall lithium content as determined by XRD fitting (red) compared to the lithium amount calculated from delivered capacity (blue) during the first charge and discharge cycles at a C/5 rate for the pristine operando cell (left) and aged operando cell 1 (right). (b) 2D lithium concentration map at the end of C/5 charging for pristine (left) and aged graphite (cell 1, right) electrodes, with the color bar indicating the x value in $Li_xC_6$ averaged along the x-axis (X-ray beam direction). (c) 2D deviation maps obtained from (b) reporting values of x-<x> in each pixel, where <x> is the average lithium content obtained in the whole electrode. Red shows an excess in Li while blue indicates local Li depletion with respect to the spatially-averaged behavior.

The 2D lithium concentration and Li concentration deviation maps were generated across a 3 mm length (abscissa) and 70 µm depth (ordinate) (Figure 6b-c and Supporting videos S1-2 (pristine) and S3-4 (aged) for complete lithiation maps). For the pristine graphite at the end of the charge (Figure 6b, left), the electrode has a color gradient showing the presence of Li concentration heterogeneity, as expected. Pixels near the separator show a larger fraction of the fully lithiated stage 1 phase (x = 1, $LiC_6$) with x~0.9, compared to the rest of the electrode (orange pixels with averaged values x~0.75 are numerous at the bottom of the electrode, for instance). This is even clearer when looking at the deviation map (Figure 6c, left) , showing accumulation (depletion) of Li at the separator (current collector) side, with respect to the average lithiation degree. This suggests that the graphite particles near the separator are subject to a higher local current density than those deeper within the electrode, even at a relatively slow cycling rate. Similar behavior has been observed in other studies, such as Finegan et al.[22], where the highest intercalation also occurred on the separator side.

In contrast, the lithiation map of the aged graphite electrode at the end of the C/5 charge (Figure 6b, right) reveals a more highly heterogeneous lithium distribution, both vertically and horizontally. Indeed, blue, green, yellow, and red regions are observed, corresponding to approximate lithiation states x = 0, 0.4, 0.6, 0.9. Clearly, a significant heterogeneity in lithium distribution is caused by aging. Additionally, contrary to the pristine electrode, the aged electrode exhibits less lithiation near the separator than near the current collector, as also clearly revealed by the deviation map (Figure 6c, right)

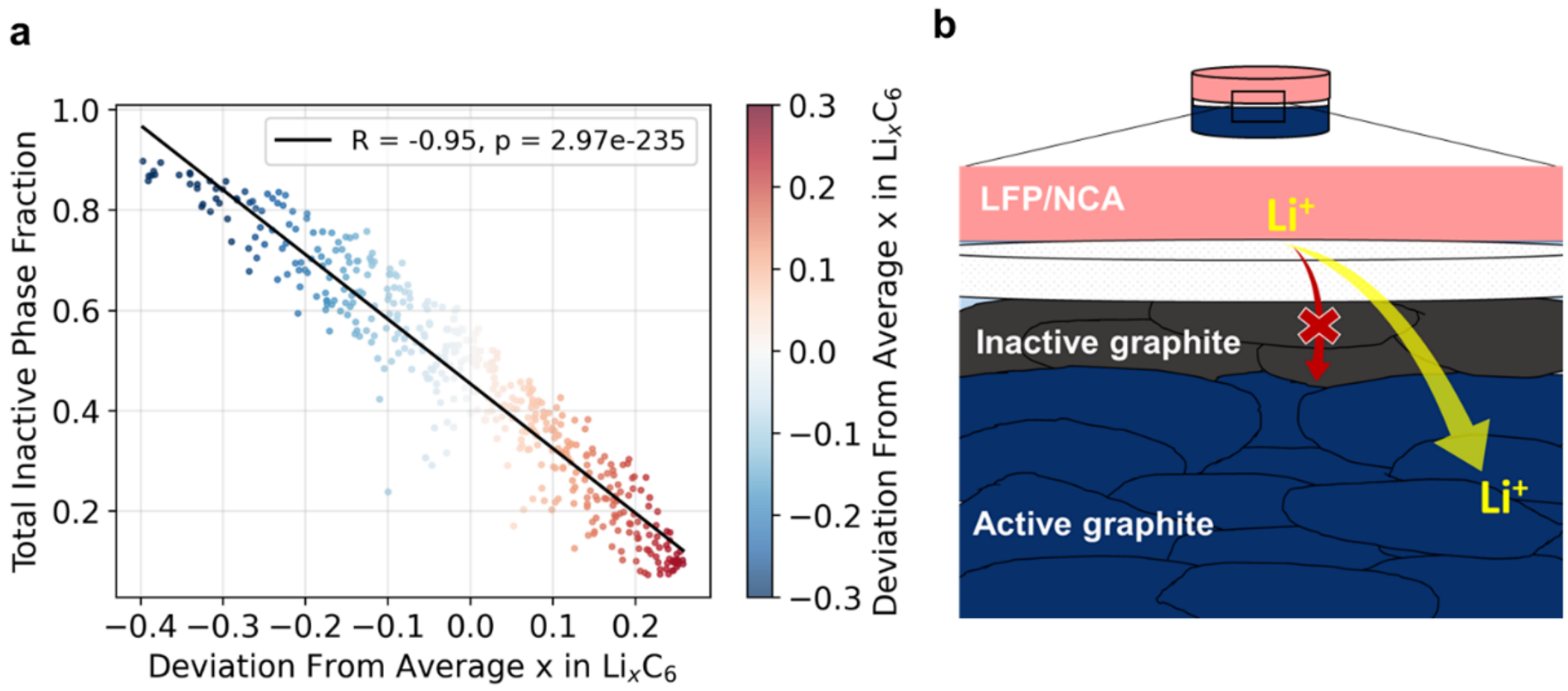

**Figure 7.** Inactivity and lithium concentration relation. (a) Correlation between the overall inactivity fraction and the Li concentration deviation at the end of charge C/5 for the aged electrode. (b) Schematic representation of the alteration of the lithiation process due to the inactivity near separator when the graphite ages.

While this behavior may seem counterintuitive, it aligns with the enhanced inactivity observed near the separator (Figure 5c) discussed earlier. The total inactivity map shows a 2D distribution of colored pixels that strikingly resembles that of the 2D lithium concentration one, with the depleted Li region matching visually with the most inactivated part, close to the separator (Figure 6b and c, right). To further investigate this phenomenon, a pixel-based correlation analysis was performed to explore the relationship between inactivity and lithiation. A linear regression model was employed, analyzing the deviation factor at the end of the C/5 charge cycle against the overall inactivity (Figure 7a). An R-value of -0.95 indicates a strong negative correlation, where greater inactivity corresponds to lower lithiation, while a p-value near 0 confirms statistical significance.

These results suggest that regions near the separator become largely inaccessible for lithiation in aged electrodes, while areas closer to the current collector exhibit greater lithiation due to the higher electrochemical activity of graphite (Figure 7b). In pristine electrodes, the highest lithiation occurs near the separator; however, in aged electrodes, this region exhibits significantly higher inactivity, suggesting a direct link between current distribution and premature aging. This degradation is likely driven by prolonged exposure to high local current densities, accelerating structural and electrochemical deterioration.

This redistribution of activity raises critical questions about how aging affects (de)lithiation mechanisms in the remaining active fraction of the electrode. To investigate this, we now turn to a quantitative analysis of lithium concentration heterogeneities and their evolution over different cycling rates.

### 2.5. Quantification of spatial heterogeneity in graphite electrochemical activity

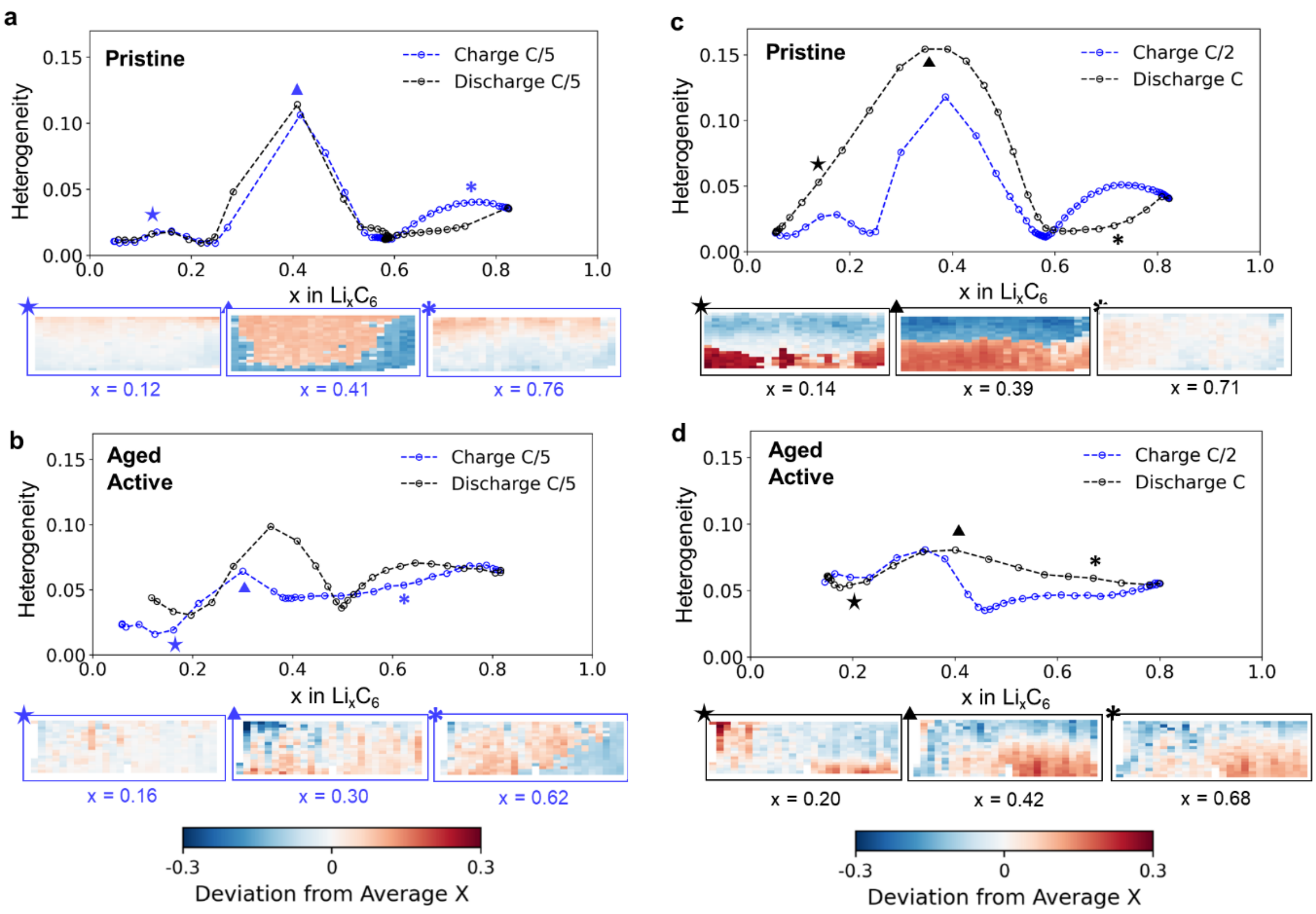

**Figure 8.** Quantifying the effect of heterogeneous aging on reaction mechanisms. (a-c) Heterogeneity factor evolution plotted for pristine electrode for (a) C/5 charge (blue) and C/5 discharge (black), for (c) C/2 charge in blue, C discharge in black. The heterogeneity factor at a given time (or SoC, x in $Li_xC_6$) corresponds to the (y,z)-integration of the deviation map at this time. Deviation maps represent the difference x - <x>, where x is the local lithium concentration and <x>, is the spatial average across the electrode. Three key states during (dis)charging are highlighted with symbols, showing the corresponding maps. (b-d) Heterogeneity evolution plotted for active graphite of aged cell 1 for (b) C/5 charge (blue) and C/5 discharge (black), for (d) C/2 charge in blue, C discharge in black. Deviation maps similarly highlight three states during (dis)charging.

To systematically compare pristine and aged electrodes across different cycling rates, we quantified lithium concentration gradients within the electrode's active fraction of graphite particles (i.e., excluding inactive graphite from the analysis). This was achieved using a heterogeneity factor, a metric that captures spatial variations in lithium distribution during electrochemical operation. Derived from 2D lithium concentration maps as a function of time and C-rates, this metric calculates at each SoC the spatially averaged heterogeneity by integrating over (y,z) the difference between the local lithium concentration x(t) at each pixel and the electrode-averaged value $<x(t)>$ (see Supporting Information for details). Therefore, it quantifies the deviation to the mean electrode behavior and varies in function of the lithiation degree[36], comprising a succession of minima and maxima that correspond to homogeneous and heterogeneous (de)lithiation, respectively. Figure 8 reports the heterogeneity factor measured in pristine at low (C/5) and higher (C/2-C) cycling rates (top panels), compared to the same metric extracted for aged electrode (bottom panels), together with selected lithium gradient concentration maps taken at key moments during the (dis)charge, visually showing the degree of heterogeneity both in-depth and laterally.

In the pristine case, three local maxima are observed while charging at a slow C-rate (C/5) (Figure 8a), corresponding to x values of 0.15, 0.4, and 0.7. Note that the heterogeneity value of the maximum at x = 0.4 has a bias coming from the uncertain quantification of Li concentration in between these x values, as discussed earlier. However, the sequence of maxima align well with phase transitions reported in the literature, respectively, to Stage 2L ↔ 2 and Stage 2 ↔ 1 for x = 0.4 and x = 0.7.[36] During discharge, only two maxima are observed (x = 0.15, 0.4), as already reported and ascribed to reduced (de)intercalation kinetics at high stoichiometry[21], responsible for a more homogeneous process for x > 0.6.

The active aged graphite cycled at low C-rate (Figure 8b) shows a reminiscent minima-maxima sequenced shape, but with significant changes as compared to pristine. First, only one maximum is observed at around x = 0.3, shifted from the x = 0.4 position seen in the pristine electrode. During discharge the maximum shifts back to x = 0.36, with a new maximum emerging at x = 0.63, hence the asymmetry between charging and discharging situations is enhanced. Moreover, a notable difference between pristine and aged is the magnitude of the maximum in the region between x = 0.6 and 0.8, which is stronger *during charge* in the pristine electrode but stronger *during discharge* in the aged electrode. This observation is robust, as it involves $LiC_6$ and $LiC_{12}$, i.e. phases with well-defined diffraction peaks and unambiguous lithium quantification. As mentioned earlier, the interplay between lithium diffusion and charge transfer was shown to govern the intensity of this maximum.[21] The difference we observe indicates that in aged electrode the phenomena driving (de)lithiation dynamics are distinctly balanced, as compared to pristine. This can be explained by the more constrained diffusion in the aged electrode (due to reduced porosity and increased tortuosity) and modified charge transfer due to surface modification of the graphite particles upon aging.

Another remarkable feature in aged electrode is the absence of the first maximum observed in the pristine electrode at low lithium content (x = 0.15). This absence might be due to blurring of the overall heterogeneity distribution, which smears out well-defined local maxima at low stoichiometry. This effect is also evident when considering values of the local minima – corresponding to regions where significant changes in the electrochemical potential drive a more homogeneous (de)lithiation process. Indeed, while the heterogeneity factor in the pristine electrode approaches zero at x = 0.25 and x = 0.6, the aged electrode shows much less pronounced minimum (close to 0.02), indicating that the active graphite in an aged electrode is *never homogenously (de)lithiated*.

Higher C-rates alter the heterogeneity profile both in pristine and aged electrodes. In the pristine (Figure 8c), the C/2 charge shows a behavior similar to that of C/5 with accentuated heterogeneities at maxima. However, discharge at C exhibits strong heterogeneity for x < 0.6, where the electrode is never homogeneous and three phases coexist at the electrode scale ($LiC_{18}$, $LiC_{12}$ and $LiC_6$). This can be interpreted as the kinetic effect kicking in, where thermodynamic equilibrium is not achieved. Consequently, Li-poor and Li-rich regions co-exist at the electrode scale. The aged electrode (Figure 8d) shows heterogeneity curves increasingly flattened (i.e., less defined maxima-minimum) as the charging rate increases, with heterogeneity values kept in a restricted range from 0.03 to 0.07. Even at the C discharge rate, the maximum heterogeneity value remains relatively low compared to the pristine electrode at the same rate, while the minimum is always greater. As the sequence of maxima and minima arise from the intrinsic shape of the electrochemical potential that alternates plateaus and sloped regions, the results indicate that the (de)lithiation process in aged electrode is spatially altered.

In conclusion, we made a series of important observations. First, there is an asymmetry between charge and discharge in the pristine and aged electrode during the Stage 2/2L to Stage1 transition. This is also observed in other characterization works but not yet understood or modelled. Second, aging causes a change in the spatial distribution of the Li-rich/Li-poor phases. On one hand, for the pristine electrode, we observed the presence of clear Li-rich/Li-poor phase boundary propagating through the depth of the electrode resulting in distinct maxima/minima in the heterogeneity factor profile. This is in agreement with previous characterization experiments and modelling results.[21,37] On the other hand, in the aged electrode, we unveil that the reaction occurs at the microscale corresponding to small domains reacting one after the other without clear depth dependance. As a consequence, the heterogeneity factor minima and maxima are blurred for the aged electrode.

### 2.6. Discussion

We have applied a detailed characterization workflow on an aged graphite electrode extracted from a commercial element to understand the nature of the aging mechanism. The workflow consisted in classical characterization methods such as electrochemical measurements, ex situ X-ray diffraction and tomography together with advanced operando µXRD imaging technique which was never been applied on aged electrodes. Combining the results from these different techniques, we have been able to observe with unprecedented spatial resolution the degradation of this graphite electrode. Our main findings are (1) the capacity of the commercial cell is mainly limited by the LCL and there is also a strong loss of active graphite (39% +/- 8%), (2) inactive graphite is found mostly at the separator surface of the electrode, (3) most of the inactive particles are either pure graphite or Stage 2/2L phase, showing the presence of Li trapping, (4) with the increase of the C-rate, we observed more inactive graphite but the spatial localization does not change, (5) spatial reaction heterogeneity in the *active* graphite particle is affected by aging, in particular, the lithiation fronts propagating from the separator to the current collector surface in the pristine electrode are not observed in the aged electrode. In the following, we compare our findings with previous characterisation work on graphite aging, starting with the presence/quantification of loss of active graphite.

The amount of active graphite loss quantified using electrochemical methods varies a lot in literature. For example, Kobayashi et al.[38] found 1% while Klett et al.[9] reported 30 - 50% in cells cycled up to 73% and 80% of capacity loss, respectively. Clearly, loss of active graphite depends on the aging conditions. Dubarry et al.[39] analyzed the capacity loss of several commercial cells and found two different capacity loss regimes with cycle number. In the first stage, the capacity loss is relatively mild and attributed to LCL, and in the second stage, capacity degradation accelerates when the loss of active material kicks in.[11,40] In other words, higher cycle number leads to more loss of active material. In light of this observation, the different loss of active graphite observed by Kobayashi et al. and Klett et al. is in line with the number of the cycles performed by these cells namely 600 and 2250 cycles. The large difference in performance between the cells in these two works originates from the cell chemistry, $LiMn_2O_4$-NCA/Gr and LFP/Gr, respectively. Indeed, $LiMn_2O_4$ is not as stable as LFP due to Mn dissolution for example. If we focus on LFP/Gr cells, which is the chemistry providing the best capacity retention, we find a rough correlation between cycle number of the amount of active graphite loss (Safari et al.[23] – 1 year aging – 10%, Klett[9] 2250 cycles – 30 – 50%), and our inactive graphite fraction is in the range of the literature for similar cells, and hence is a relevant aging mechanism which needs to be better characterized.

Electrochemical methods can not inform us on the nature and the spatial localization of the inactive graphite phases. In particular, it is important to understand if Li is trapped in lithiated inactive graphite particles, and where it is located in the electrode. Rauhala et al.[41] could visually observe regions with bluish color at the surface of aged graphite electrodes suggesting the presence of electrochemically inactive Stage 3L at the discharged state. However, the authors did not quantify the amount of trapped Li. Mikheenkova et al.[13] performed *ex situ* XRD in plane mapping with a 1 mm² beam of an aged graphite electrode and found Li trapped into inactive parts of the electrode (Stage 4 or Stage 3L). Both authors found that the inactive graphite is heterogeneously dispersed in the electrode plane but give no information how the inactive particles are distributed in the electrode depth. Scipioni et al.[17] found graphite particle disconnected from the electronic network at the separator side of the electrode using SEM. However, electronic disconnection is not the only source of particle inactivity (pore clogging, structural degradation), hence the need for more information.

In our system, for the first time, we determined the nature of the inactive phases and spatially localized these phases in the electrode depth. We observe up to 11% of inactive Stage 2/2L, hence a larger amount of trapped Li compared to Mikheenkova report. This could be due to differences in the aging conditions since Mikheenkova's NMC/Gr cells reached end of life after only approx. 600 cycles. Moreover, inactive pure graphite is found at the separator side of the electrode, while lithitated inactive graphite phase are buried slightly underneath this inactive graphite layer. This is in line with the surface deactivation observations from Rauhala et al.[41] or Scipioni et al.[17] We hypothesize that separator-side inactivation results from "overworked" graphite. Indeed, the separator side of the electrode experienced more complete (de)lithiation cycles in pristine conditions compared to the current collector side, an effect that might lead to premature aging on the separator side.

The premature aging could be accelerated by Li plating occurring at the surface of the electrode due to the accumulation of $LiC_6$ saturating all the Li intercalation sites as hypothesized by Weisenber et al.[42] The presence of $LiC_6$ at the surface for pristine electrode has been demonstrated by several works including our team.[21] However, prior to our work, there was no operando experiments quantifying Li concentration heterogeneity in an aged electrode depth. As a matter a fact, in our aged electrode, we do not observe a saturation of the active surface by $LiC_6$ phase. Instead, we find that the $LiC_6$ active phase is scattered in the electrode depth. This is due to a switch of the reaction mechanism heterogeneity from the macroscale (electrode level) to the microscale (group of particle) with aging. The physical parameters that define whether the reaction mechanism occurs at the macro or micro scale are: (1) the homogeneity of tortuosity—at a given depth in the electrode, how equally accessible are the graphite particles? (2) the homogeneity of charge transfer—at a given depth, how equally lithiable are the particles? and (3) the homogeneity of Li solid diffusion within the particles—at a given depth, how equally does Li diffuse within individual particles? The pristine electrode clearly corresponds to the first scenario (macroscale heterogeneity) but during aging, processes such as pore clogging, deactivation of graphite particles, particle surface modifications, and higher lithiation of particles near the separator introduce heterogeneities. Therefore, the uneven distribution of dead and slow graphite particles induces a large variability in the local concentration of lithiated phases, which entails local disparities in overpotentials and ion transport pathways. As a result, aging shifts the reaction mechanism toward the microscale scenario in the active fraction of the electrode, resulting in profoundly altered charge dynamics in aged negative electrode.

## 3. Conclusions

In this work, we investigated the aging-induced structural and morphological changes in graphite electrodes extracted from a large-format cell cycled to 70% of remaining capacity. Ex situ chemical, structural, and morphological analyses confirmed the coexistence of two primary aging mechanisms: active material loss and cyclable lithium loss. Using operando µXRD we revealed how aging fundamentally alters reaction mechanisms in graphite electrodes, marking the first time application of this technique to aged graphite electrodes.

Inactive graphite consists of both unlithiated and lithiated phases, suggesting that some particles retain lithium. Inactivity is C-rate dependent, and faster cycling generates more lithiated inactive particles, indicating the existence of two types of aged graphite: dead (i.e., permanently inactive and electrochemically lost) and slow (i.e., kinetically hindered but potentially reactivatable at lower cycling rates). Globally, a significant fraction of active material is lost, predominantly near the separator, where higher local current densities accelerate degradation. Aging disrupts lithiation pathways, shifting from a well-defined reaction front at the macroscale in early cyclings to a heterogeneous coexistence of multiple phases at the microscale in aged electrodes. This results in persistent spatial heterogeneities, even in regions where graphite remains electrochemically active.

While inactive active material during aging has been reported previously for other chemistries, this study provides a quantitative assessment of the amount, phases, and spatial localization of inactive graphite. A notable hypothesis emerging from our findings is that separator-side inactivation results from “overworked” graphite, which experienced more complete (de)lithiation cycles in pristine conditions compared to the current collector side, an effect that might lead to premature aging on the separator side.

Our findings suggest that aging is a localized, phase-dependent transformation that shapes the electrochemical and structural evolution over extended cycling. Understanding these degradation pathways is crucial for graphite electrode design and potential strategies to mitigate performance decay, including modifying electrode architecture, optimizing electrolyte formulations, and tailoring cycling protocols to enhance long-term stability.

## 4. Experimental Section

### Material preparation

All electrodes used in this study were manufactured for research purposes. The aged positive and negative electrodes had been cycled in a commercial cylindrical cell several thousand times until they reached 70% of their initial capacity. The cycling procedure consisted of charging the cell at C/2 to 3.6 V, then at C/5 to 3.7 V, followed by C/11 and C/45 to 3.8 V. After charging, the cell rested for 15 minutes before discharging at C/2 down to 2.5 V, followed by another 15-minute rest. This cycle was repeated four times, followed by a 24-hour resting period. Every 24 cycles, a control cycle was performed: charging at C/3 to 3.6 V, then at C/5 to 3.7 V, followed by C/11 and C/45 to 3.8 V, with a 5-minute rest after charging. The cell was then discharged at C/3 to 2.5 V, followed by another 5-minute rest. All tests were conducted at room temperature. Both pristine and aged electrodes were comprised of an 80 µm-thick positive electrode, cast on a 15 µm-thick aluminum foil, which contained a mixture of $LiFePO_4$ /Li(NiCoAl)$O_2$ (in a 90/10 wt% ratio) as the active material. The negative electrode, 70 µm-thick, contained graphite (Gr) as the active material and was cast onto a 10 µm copper current collector. Disks measuring 1.6 cm in diameter of the pristine electrodes were assembled in a CR-2032-type coin cell in full cell configuration within an Ar-filled glovebox. A polypropylene-based separator and 50 µl of $LiPF_6$ salt-containing carbonate-based electrolyte were used in the assembly process. The prepared coin cells underwent two cycles at C/10 (corresponding to a full charge in 10 hours) at 60°C, followed by one cycle at C/5 at room temperature to establish the initial solid electrolyte interphase (SEI) layer. The lower and upper cutoff voltages for the formation cycles were 3.8 V and 2.5 V, respectively. To ensure the reliability of the results, multiple coin cells were cycled using both pristine-formed and extracted aged electrodes. At a C/5 cycling rate, the average discharge capacities at room temperature were $Q_{disc,aged} = 1.66 \pm 0.32$ mAh/cm² for aged full cells and $Q_{disc,pristine} = 2.26 \pm 0.14$ mAh/cm² for 25 pristine full cells.

### Cycling conditions

For operando experiments, the positive and negative electrodes from formed pristine and as-received aged electrodes were punched into 3 mm diameter disks and mounted on a swagelock-type operando cell, which has a perfluoroalkoxy alkanes (PFA) casing in an Ar-filled glovebox. A polypropylene-based separator and 2.5 µl of $LiPF_6$ salt-containing carbonate-based electrolyte were used. A pristine (i.e. uncycled) polypropylene-based separator is used for both pristine and aged cells. To check the assembled cell quality before the operando experiment, all prepared operando cells underwent one cycle (comprising a complete charge and discharge) at a C/5 rate. The prepared pristine and as-received aged electrodes were cycled within the operando cells following a standardized protocol at room temperature using a Biologic potentiostat. This protocol involved a complete cycle (charge and discharge) at C/5, followed by a subsequent cycle at C/2 cycling rate, then one charge at C/2, and a faster discharge at C, all within the potential range of [2.5 to 3.8 V]. Each charge and discharge cycle was followed by a 1-hour hold at 3.8 V and 2.5 V, respectively, to ensure a complete (de)lithiation. To ensure the representativity of the electrochemical data obtained with the operando cell, it was compared with data from coin cells assembled with the same separator and electrolyte for both pristine and aged samples.

**Nanoholotomography experiment**

Nanoscale holotomography[33] was performed at the ID16A beamline at the European Synchrotron Radiation Facility (ESRF). 1 mm of pristine and aged graphite electrodes were assembled in a glass tube, mounted on a pin adapted for tomographic experiments, and placed on a sample tray. The sample was positioned on a motorized stage between the focal point and the detector, and acquisitions were conducted using holotomography and phase contrast imaging. Four tomographic scans were acquired at different distances between the focal plane and detector. Each tomographic scan included 2000 projections acquired over 180° with an exposure time of 0.150 s per projection, leading to a total acquisition time of ~3 hours 30 minutes per electrode.

3D reconstructions were performed in two steps. First, we have performed phase retrieval using an in-house Octave script. The procedure starts with a Paganin-like approach[43], with a $\delta/\beta$ ratio of ~2800, to retrieve an estimate of the phase distribution. This is followed by a nonlinear conjugate gradient optimization. Second, the projections of the phase are used to reconstruct the volume using filtered back projection algorithm implemented in Nabu software package. The phase contrast in reconstructed volumes is related to the electron density variations in the material. The final 3D volumes were reconstructed with a 25 nm voxel size in 32-bit floating point format. Image analysis was carried out on the full reconstructed volume using ImageJ and Python to evaluate porosity, while 3D visualizations of the electrode were generated using ParaView®.

**Laboratory X-ray diffraction**

Ex situ X-ray diffraction (XRD) measurements were performed using a Bruker D8 Discover diffractometer with Cu $K\alpha_1$ radiation ($\lambda$ = 1.5406 Å) in a 2θ configuration, ranging from 15° to 80° with an angular resolution of 0.01°. XRD measurements were conducted on electrode pieces extracted from the large cycling element or on electrodes reassembled in coin cells. These samples were re-mounted alone in an air-tight operando cell[44] to prevent exposure to ambient conditions. The XRD patterns were fitted using the WinPlotr[45] program, employing pseudo-Voigt functions with a globally defined full width at half maximum (FWHM) and eta (the proportion of the Lorentzian component). A linear background was applied during fitting. Each Bragg peak was characterized by its position, intensity, FWHM, and eta shifts relative to the global parameters.

Wide-angle X-ray scattering (WAXS) measurements were performed on a larger piece of the aged electrode sheet using a Xenocs Xeuss 3.0 instrument equipped with a copper source (Cu $K\alpha_1$ radiation, $\lambda$ = 1.5406 Å) operating in transmission mode. The scattered patterns were collected using a Dectris Eiger 2R 500K detector. The sample was placed 55 mm before the detector on a motorized stage within a sealed aluminum pouch to prevent air exposure. Scans were conducted across a Q-range of 0.1 to 3.4 $Å^{-1}$, with a beam spot size of 900 µm, covering a 3 x 4 $mm^2$ region. The diffraction rings were azimuthally integrated[34], and the X-ray diffraction data were processed, background-subtracted, and fitted using the Prisma tool[35].

**Operando micro X-ray diffraction experiments**

Operando µXRD mapping was conducted at the European Synchrotron Radiation Facility (ES RF) on beamlines ID13 (cell 1) and ID31 (cell 2).

For the ID13 experiment, a focused 3 x 3 µm beam (in the vertical and horizontal directions, z and y) at an energy of 23.5 keV was used. The battery stack was placed inside a 3 mm diameter Swagelok type cell. The stack was composed of the LFP/NCA positive electrode, graphite negative electrode, polypropylene based separator and 2 µL carbonate based electrolyte (same compounds as used in the coin cell tests described above). The battery was positioned so that the X-ray beam passed through it along the x-axis. During data collection, the battery was moved horizontally (y) and vertically (z) to scan the entire electrode stack, producing (y, z) maps with 35 x 28 pixels, with each pixel measuring 100 x 3 µm, in approximately 2 minutes. Each position was counted for 0.01 seconds. "Horizontal fly scans" were performed by continuously moving the battery along the y direction from -1.6 to 1.6 mm, while the shutter remained open. Detector images were averaged over 100 µm of battery displacement, meaning each image was an average of 100 µm of the sample in the y direction. After each horizontal continuous scan, the battery was moved vertically by 3 µm (along the z direction), and a new continuous scan in the y direction was performed until the complete 2D maps were obtained. This continuous scanning method minimizes dose exposure by spreading it over 100 µm of horizontal displacement. WAXS patterns were recorded using a 2D CdTe detector in the Q range of 1.18 – 3.66 $Å^{-1}$. The detector calibration was performed using a reference sample ($Al_2O_3$) with PyFAI[46], and the sample-to-detector distance calibration was corrected based on the Cu current collector peaks.

For ID31, the experiment is very similar to the ID13 experiment with however some differences which are listed here. The energy is 75 keV, the beam size is 20 x 8 µm (y, z), number of scanning points is 37 × 13 pixels (y, z) over 3.6 mm and 90 µm (y, z directions). WAXS patterns are collected on a 2D Pilatus CdTe detector of 2M pixels, collecting the full diffraction rings due to the larger size of the detector. Q Range is 0.47 – 6.29 $Å^{-1}$. The detector calibration was performed using a reference sample ($CeO_2$) with PyFAI[46], and the sample-to-detector distance calibration was corrected based on the Cu current collector peaks.

## Acknowledgments

This work is part of a joint TotalEnergies and CEA project. The authors acknowledge the European Synchrotron Radiation Facility Extremely Brilliant Source (ESRF-EBS), Grenoble, France for providing experimental time (proposals numbers: IN-1150 for ID13 and Battery Pilot Hub MA-4929 “Multi-scale Multi-techniques investigations of Li-ion batteries: towards a European Battery Hub” for ID31 and ID16A) and thank the staff of beamlines ID13, ID31 and ID16A-NI at the ESRF-EBS for their assistance and support during the experiments. The authors thank Sandrine Schlutig at CEA/IRIG/MEM for technical support on the lab XRD and helpful discussion.

## Data Availability Statement

The data supporting the findings are available within the main text and ESI. More detailed data, as well as the code to process the raw data can be made available upon request.

## Authors Contributions

Q.A. and S.L. coordinated the project. S.L., S.T., and Q.J. conceived the synchrotron microdiffraction investigation and performed the micro XRD experiments on ID13 and ID31 beamlines with F.M. Micro XRD data were analyzed by F.M. and G.O. with guidance from Q.J., S.T., and S.L.. S. M. performed the holotomography experiment on ID16A and analyzed the images, discussing with I.J.. Beamlines were set up by D. K., M. M., J. D., A. M., and M. B., who helped with experimental conditions and data acquisitions. Electrochemical analysis was performed by F.M. and G.O., and the lab XRD by G.O.. All authors interpreted the combined results of the study. G.O. wrote the first draft with S.T., Q.J., Q.A., and S.L. All authors further revised the manuscript.

## Conflict of Interest

The authors declare no competing interests.

**SUPPORTING INFORMATION**

# Dead, Slow and Overworked Graphite: Operando X-ray Microdiffraction Mapping of Aged Electrodes

G. Oney[a], F. Monaco[a], S. Mitra[a], A. Medjahed[b], M. Burghammer[b], D. Karpov[b,e],

M. Mirolo[b], J. Drnec[b], I.C. Jolivet[c], Q. Arnoux[d], S. Tardif[e], Q. Jacquet[a,*], S. Lyonnard[a,*]

[a]*Univ. Grenoble, Alpes, CEA, CNRS, Grenoble INP, IRIG, SyMMES, F-38000 Grenoble*
[b]*European Synchrotron Radiation Facility, F-38043 Grenoble Cedex, France*
[c]*TotalEnergies OneTech, CSTJF, avenue Larribau, 64018 Pau Cedex, France*
[d]*TotalEnergies OneTech, Centre de recherche de Solaize, 69360 Solaize, France*
[e]*Univ. Grenoble Alpes, CEA, IRIG, MEM, 38000 Grenoble, France*

[*]*Corresponding authors:*

quentin.jacquet@cea.fr
sandrine.lyonnard@cea.fr

## Supporting Section 1: Sample integrity during electrode preparation

The extracted aged graphite electrodes were rinsed with dimethyl carbonate (DMC). DMC has low solubility for most SEI components,[1,2] and this mild washing step is therefore not expected to significantly alter the SEI layer. The extracted aged graphite electrodes were rinsed with dimethyl carbonate (DMC). DMC has low solubility for most SEI components,[1,2] and this mild washing step is therefore not expected to significantly alter the SEI layer. We checked that SEI is not dissolved by the washing procedure by performing a single electrochemical cycle prior to operando X-ray measurements. No notable reformation of the SEI was detected. For the initial cycles of aged electrodes in coin cell the calculated coulombic efficiencies are 96 ± 2%. The slightly lower coulombic efficiency with respect to the fresh cell (98 ± 1% at C/5 cycling rate after formation cycles for pristine electrodes) could rise from the use of a fresh electrolyte, and/or the increased SEI heterogeneity of the aged sample.

The potential for material detachment during the preparation of aged graphite electrodes, particularly at the cut edges, is a recognized concern. To minimize such effects, particular care was taken during sample preparation to preserve electrode integrity. Both pristine and aged were handled under inert conditions within an argon-filled glovebox, and electrode discs were punched using identical procedures.

Several lines of evidence support the reliability of the preparation method and indicate that surface and structural properties were not significantly impacted by handling.

High-resolution X-ray nano-tomography confirmed the preservation of cross-sectional morphology and the expected electrode thickness at the nanoscale. Both pristine and aged samples displayed well-maintained structural integrity, with no evidence of delamination or morphological artifacts arising from preparation.

Electrochemical reproducibility further supports the consistency of the handling protocol. For pristine electrodes, discharge capacity measurements across multiple cells showed low variability:

• Coin cells: 2.26 ± 0.14 mAh/cm² (number of cells n = 25; ~6% variability), coulombic efficiency: 98 ± 1%

• Swagelok-type cells: 2.22 ± 0.11 mAh/cm² (n = 5; ~5% variability)

This level of consistency suggests minimal disruption during preparation. In the case of aged electrodes, the observed variability was higher:

• Coin cells: 1.66 ± 0.32 mAh/cm² (n = 6; ~19% variability); coulombic efficiency: 96 ± 2%

• Swagelok-type cells: 1.23 ± 0.08 mAh/cm² (n = 5; ~6% variability)

The increased spread in aged coin cells is attributed to the intrinsic spatial heterogeneity introduced by electrode degradation. Since coin cells sample a larger surface area than Swagelok-type cells, the capacity values can vary depending on the degree of local degradation at the punched region. This spatial variability was independently confirmed by micro-X-ray diffraction (µXRD) analysis of a second electrode, which showed elevated levels of inactive graphite in certain regions.

Further spatially resolved µXRD measurements revealed no inactive graphite regions on the separator-facing side of pristine electrodes. In contrast, aged electrodes exhibited clearly defined areas of lithiated but inactive graphite, which are not consistent with artifacts from sample handling. Instead, such features are indicative of degradation mechanisms arising from electrochemical cycling. The similarity in spatial patterns observed across independently prepared aged electrodes provides additional confidence in the robustness of the preparation method and the reliability of the analytical approach.

## Supporting Section 2: Ex situ analyses

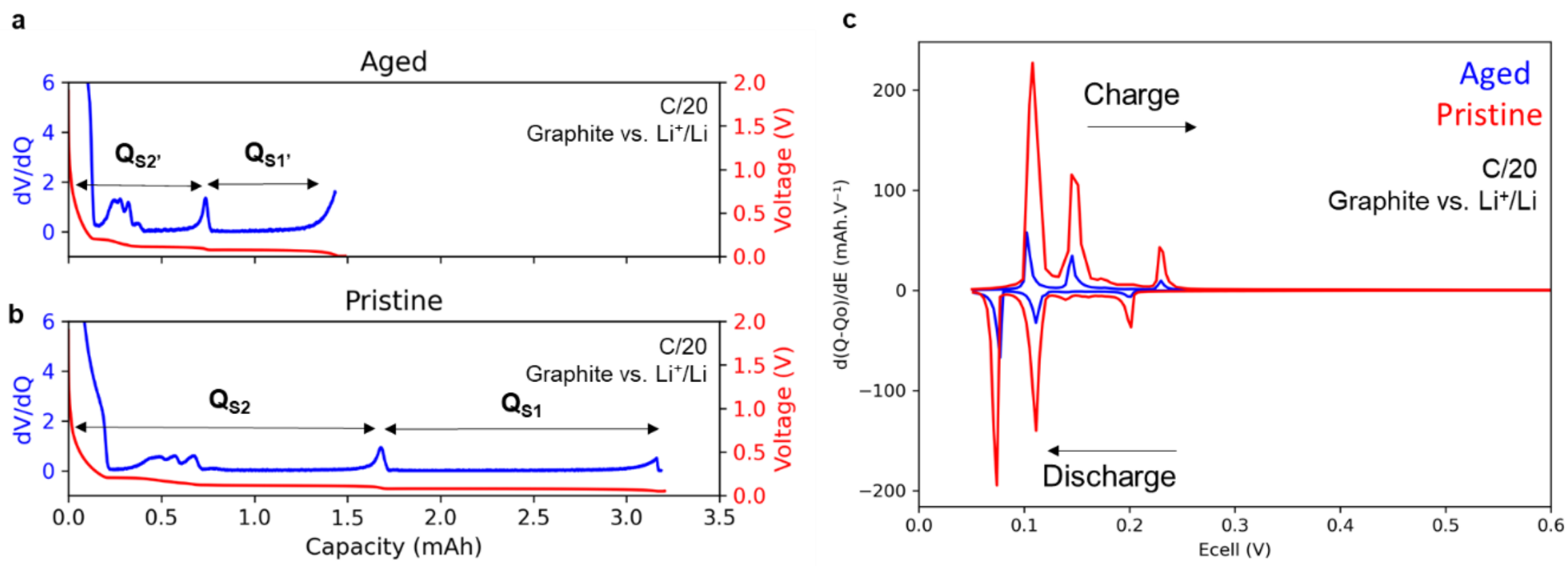

**Figure S1** Derivative voltage analysis of aged (a) and pristine formed (b) graphite half-cells (blue). The corresponding discharge profiles at C/20 within the potential range of [2-0.05 V] are shown in red. (c) Incremental capacity analysis of the charge-discharge cycle at C/20 for pristine formed (red) and aged (blue) cells.

The capacities delivered at the plateaus corresponding to the graphite transitions from unlithiated graphite to Stage 2 and Stage 2 to Stage 1 are denoted as $Q_{S2}$ and $Q_{S1}$, respectively. The values of the pristine formed graphite are 1.68 and 1.52 mAh.cm$^{-2}$, respectively. In contrast, for the aged graphite, the corresponding capacities are $Q_{S2'} = 0.74$ and $Q_{S1'} = 0.70$ mAh.cm$^{-2}$. The shrinkage in the dV/dQ plateau compared to the pristine one indicates an important loss of active material in the graphite. Since the lithium source is effectively infinite when facing lithium metal, we cannot assess the loss of cyclable lithium in this setup. The dQ/dV profile shows a significant intensity reduction for the aged graphite at each phase transition, consistent with the active material loss observed previously.

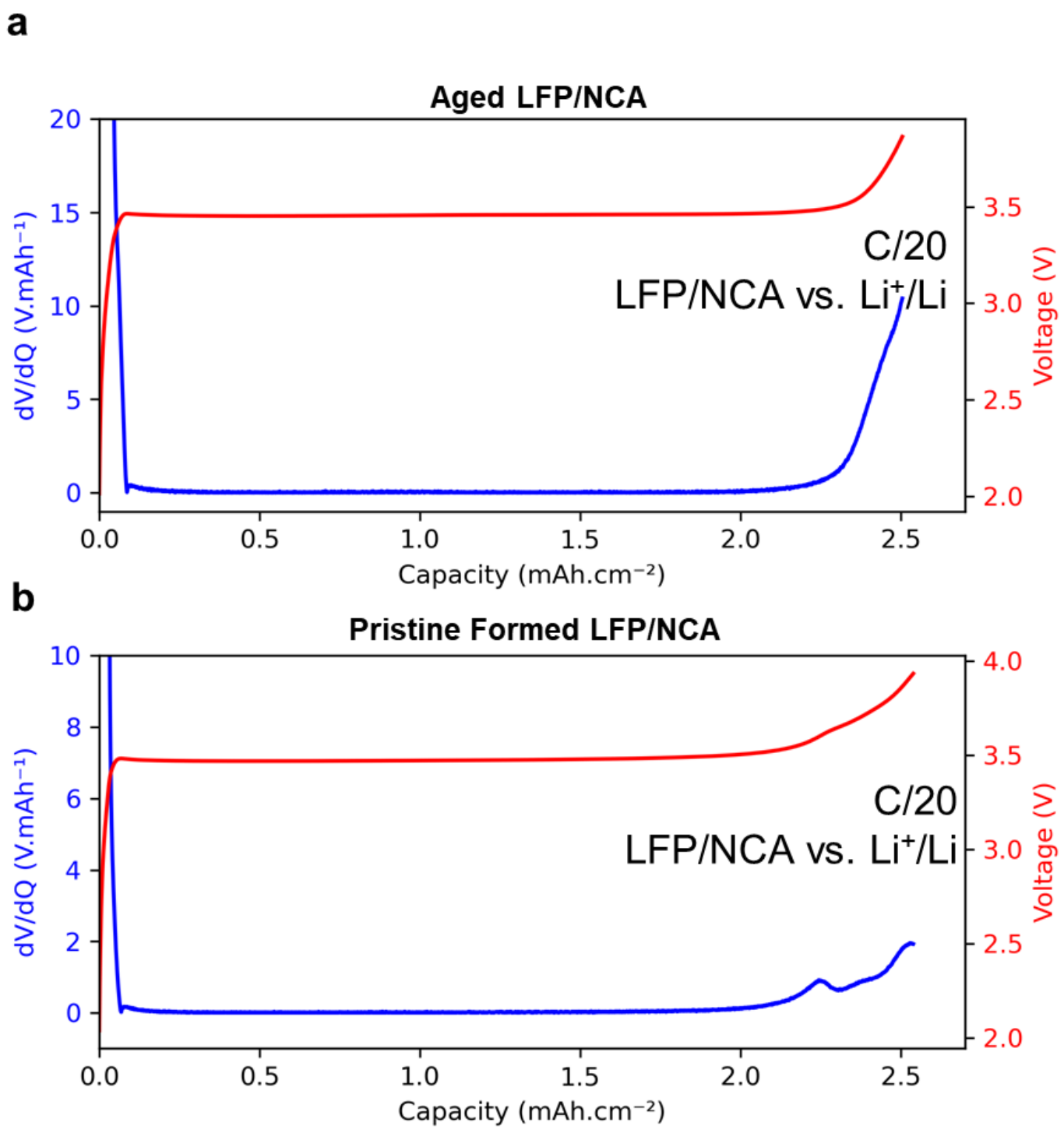

**Figure S2** Derivative voltage analysis of aged (a) and pristine formed (b) half-cells of LFP/NCA positive electrode (blue). The corresponding discharge profiles at C/20 within the potential range of [2-4 V] are shown in red.

The aged positive electrode delivers a similar capacity as the pristine formed one with 2.5 $mAh.cm^{-2}$ at the end of charge to 4V, confirming the electrochemical stability of the LFP after long-term cycling. Changes in the DVA profile at high voltage range are attributed to active material loss of NCA when the positive electrode is aged. Note that 4V is a higher cut-off voltage compared to 3.8V for full-cell configuration, hence less contribution is expected from the NCA active material.

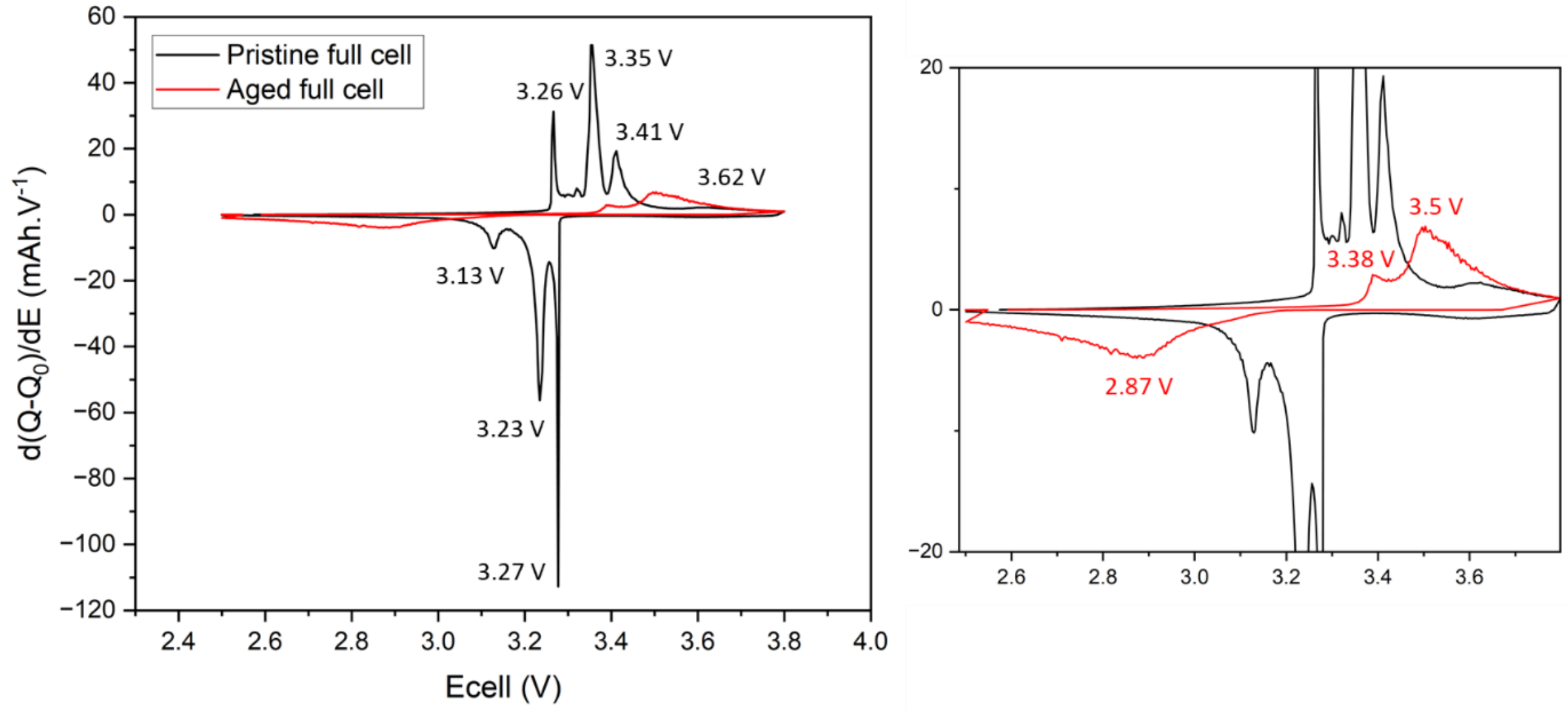

**Figure S3** Incremental capacity analysis of the charge-discharge cycle at C/5 for pristine (black) and aged (red) full cells. The right panel shows a zoomed-in view of the curves.

A significant reduction in peak areas is observed in the aged cell, indicating a loss of active material. Additionally, the noticeable shift in the main peaks reflects increased cell polarization due to aging, related to SEI-formation with loss of cyclable lithium. The NCA-related feature seen around 3.6 V in the pristine cell is no longer visible in the aged one, which we attribute to polarization effects. However, it should be noted that NCA constitutes only ~10% of the positive electrode blend, and thus its contribution to the full-cell response is relatively limited.

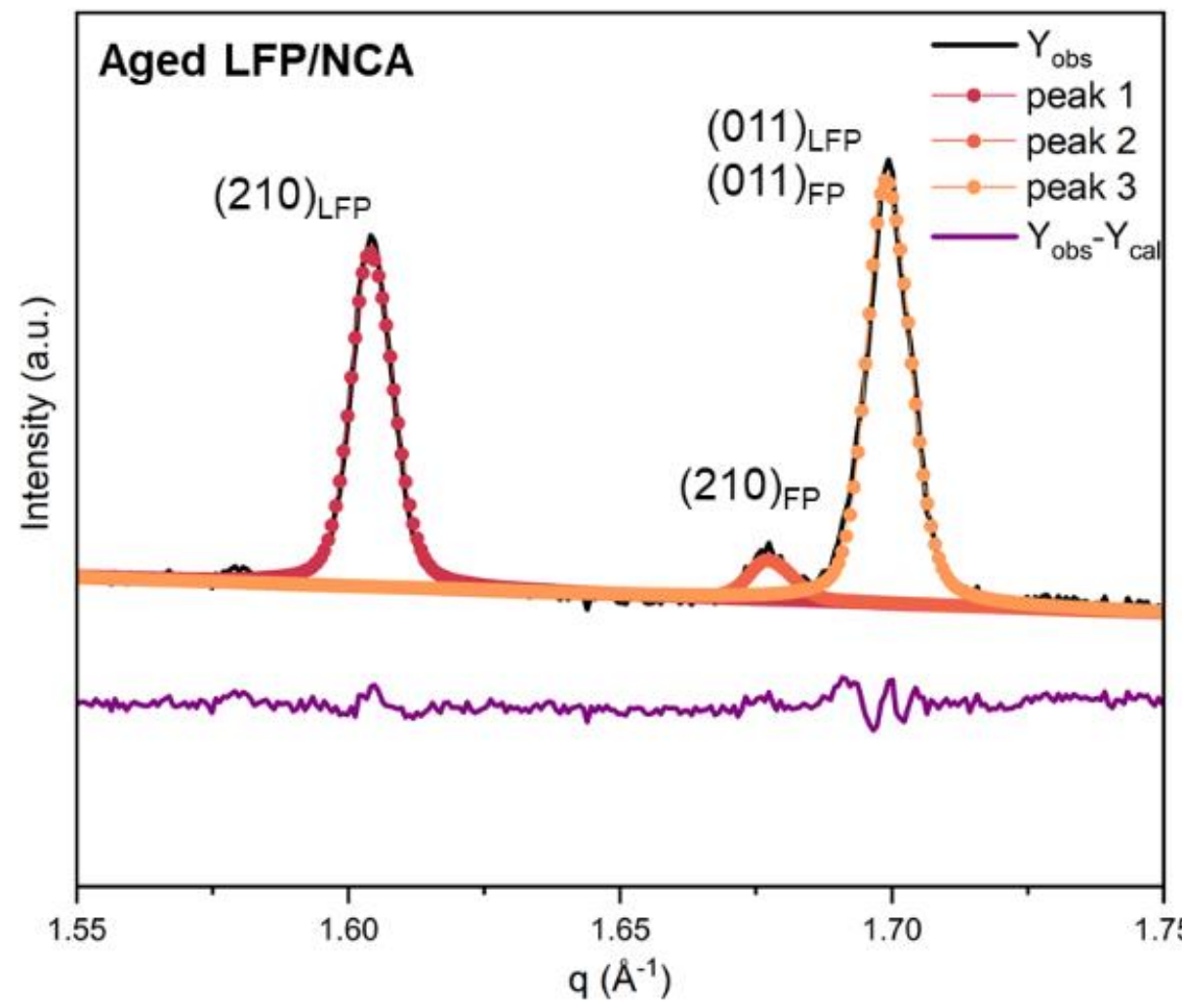

**Figure S4** X-ray diffraction pattern fit of the aged LFP/NCA electrode extracted from the cycling element at the end of life at 1.5 V, where the positive electrode is expected to be lithiated.

Profile fitting in the WinPlotr software as described in the Experimental Section. The area under each Bragg peak is calculated thanks to fitted parameters using the equation:

$$Area = h\,.\,\Gamma\,(\eta.\frac{\pi}{2} + (1-\eta)\,\sqrt{\frac{\pi}{4\ln(2)}}) \quad (1)$$

With $h$ being the peak height, $\Gamma$ being the peak width, and $\eta$ the Gaussian ratio. The ratio of the (210) peak originating from lithiated $LiFePO_4$ to the (210) peak from fully delithiated $FePO_4$ can be used to determine the $LiFePO_4/FePO_4$ ratio in the electrode. Since the structure factor for the (210) reflection is identical in both phases, the peak area ratio is a reliable approximation for the phase fraction. Analysis of the peak area ratios indicates a 60.3/39.7% $LiFePO_4$ to $FePO_4$ ratio, meaning only ≃ 60% of the LFP active material contributes to the electrochemical operation. This confirms the negative electrode slippage due to Li inventory loss.

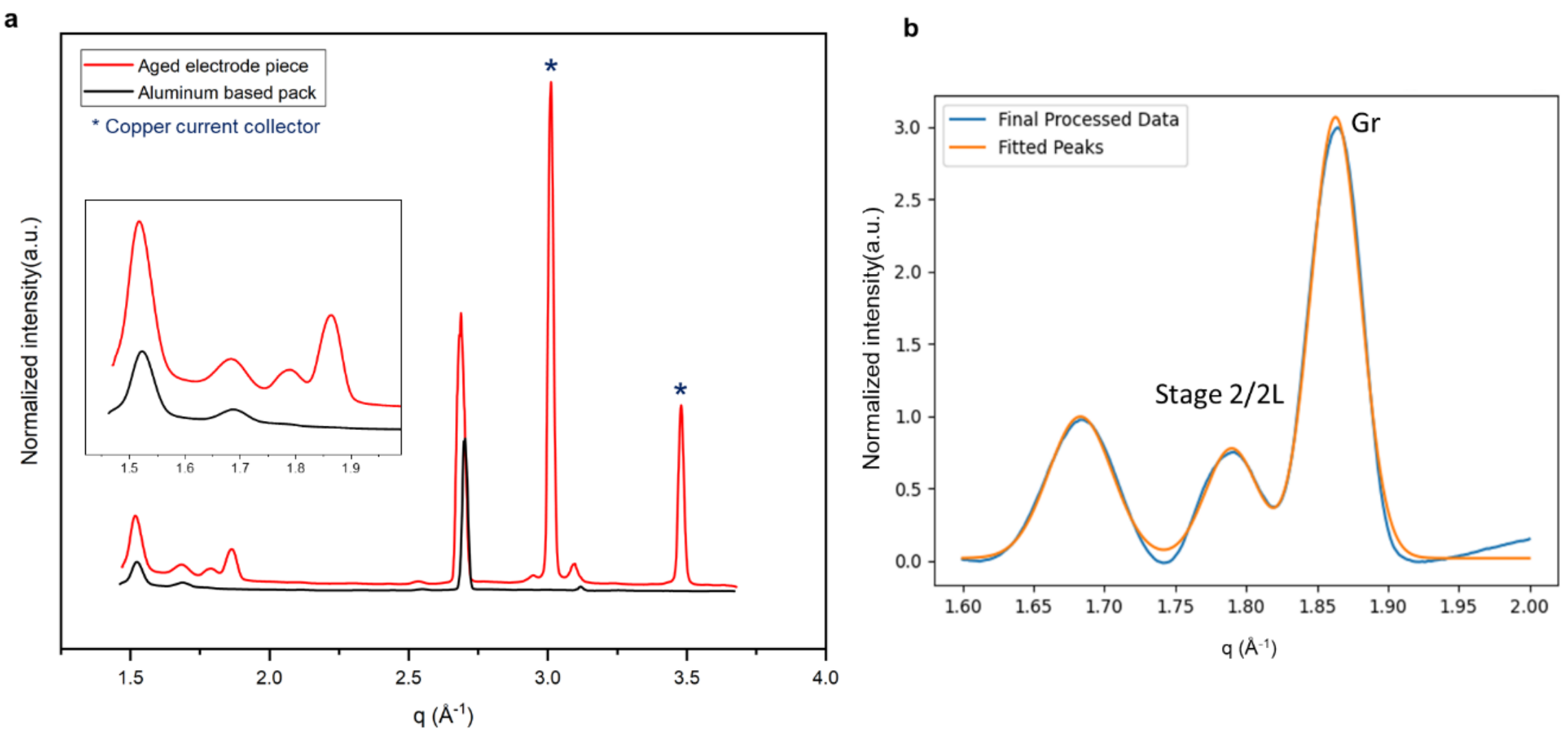

**Figure S5** (a) Averaged lab X-ray diffraction (XRD) pattern measured in transmission geometry of the aged graphite electrode extracted from the cylindrical cycling element (red). The pattern corresponds to a summed 3 × 4 mm² region. The sample was sealed in an aluminum pack to prevent air exposure during measurement. The background signal from the aluminum pack is shown in black. The asterisk marks reflections from the (111) and (200) planes of Cu foil. The inset highlights the region of interest for graphite lithiation stages. (b) Corresponding background-subtracted XRD data (blue), with profile fitting (orange) applied.

Profile fitting was performed using the PRISMA tool[1] in Python. For phase analysis, only the peaks at 1.79 and 1.88 Å were considered. Although the 1.69 Å region could indicate the presence of phase 1, aluminum background interference prevents its detection. The ratio of Stage 2-2L to graphite (Gr) is 20.8% / 80.2% by analyzing the area under diffraction peaks. This indicates that the discharging in the cylindrical element did not fully delithiate the graphite electrode due to aging effects.

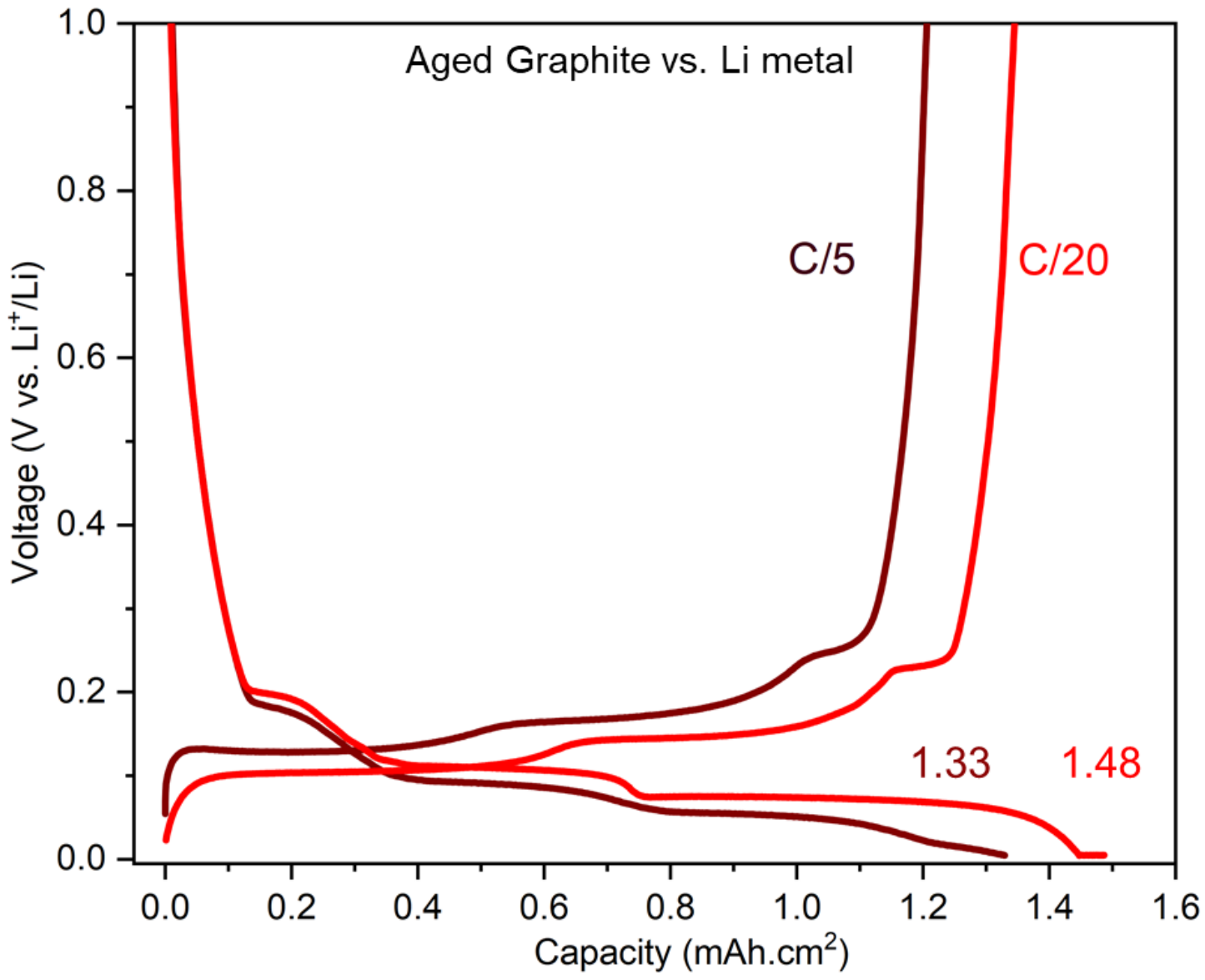

**Figure S6** Comparison of the electrochemical performance of aged graphite at different C-rates. The dark red curve represents a charge-discharge cycle performed between [0.05–2 V] at a C/5 rate, while the light red curve corresponds to the subsequent cycle at a C/20 rate.

The aged graphite electrode that is extracted from the element is re-assembled in half cell configuration. The discharge capacity observed at C/5 is 1.33 mAh.g$^{-1}$, which increases to 1.48 mAh.g$^{-1}$ when the cell is cycled at a slower rate of C/20. This confirms the presence of a kinetic limitation with graphite aging.

**Supporting Section 3: Operando electrochemistry data analysis and XRD data fitting**

**Masking acquired data.** The overall intensity of the XRD signal is first summed across all identified phases to produce a total intensity map for each spatial point and measurement. This total intensity is then normalized on a per-map basis by scaling each 2D image so that its minimum pixel value becomes 0 and its maximum becomes 1, effectively rescaling all maps to a common [0, 1] range for consistent comparison.

After normalization, a series of masks are applied to exclude low-signal or unreliable regions: pixels are retained only if their normalization factor falls within a defined range (above a minimum threshold and below a maximum one), and if the normalized intensity itself is higher than a specified intensity threshold. This masking ensures that only physically meaningful, high-quality signal regions are considered for further analysis, while edge artifacts or weak/noisy zones are excluded from the region of interest. But this reduces the field of analysis from full thickness of the elextrode (70 µm) to a restricted zone.

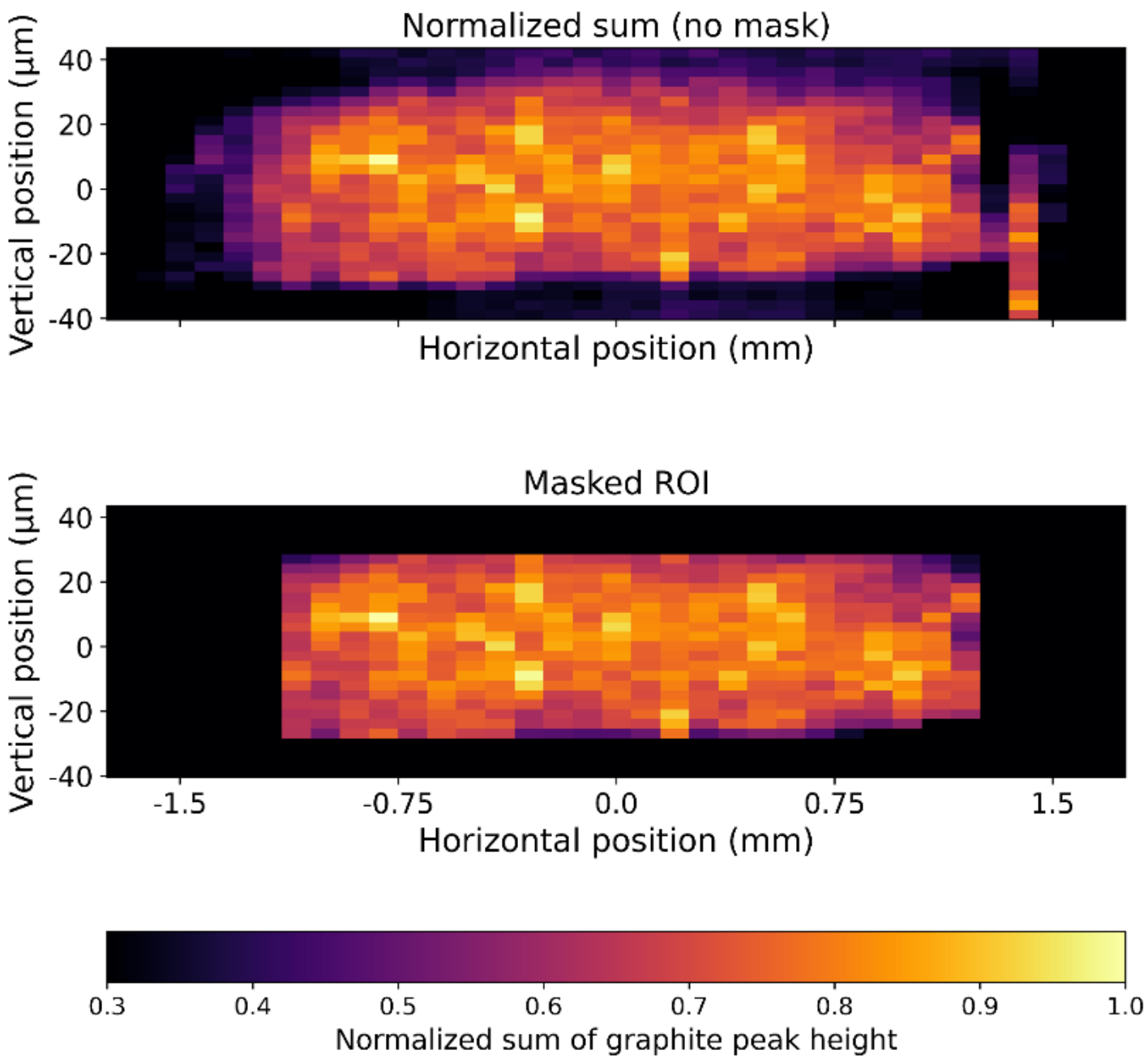

**Figure S7.** (Top) Spatial distribution of graphite within the mapped area, based on normalized diffraction peak intensities (1.7 and 1.9 Å⁻¹) (Bottom) Masked graphite distribution based on the normalized summed intensity of graphite peaks.

**Operando electrochemistry data.** Calculating x in $Li_xC_6$ during operation involves using current and capacity normalized by the electrode surface area recorded through the potentiostat. The electrochemical data points closest in time to the XRD measurements are identified to align these electrochemical measurements with XRD data. The change in Δx during each XRD map interval is determined by integrating the electrochemical current throughout the map. Mathematically, this is expressed as:

$$\Delta x = \frac{I_{avg} . t_{map}}{Q_{th}} \tag{2}$$

Where $I_{avg}$ is the average current during the interval (normalized to the electrode area), $t_{map}$ is the map duration, and $Q_{th}$ the theoretical (expected) capacity of graphite for full lithiation. The cumulative x values can be calculated by summing these contributions over time and introducing an initial $x_o$ value deducted from the X-ray diffraction pattern.

**The pattern fitting** was performed using the PRISMA tool[3] in Python. For the raw XRD patterns recovered from each pixel, the region of interest was trimmed to the q-range between 1.6 and 2 Å, and the baseline was removed. Peak bounds were assigned as follows: q=(1.65, 1.76), (1.761, 1.804), (1.805, 1.837), and (1.838, 1.9) Å , corresponding to Stage 1, Stage 2, Stage 3, and Stage 4 - graphite, respectively. A pseudo-voigt profile was selected for peak fitting, with examples provided in Figure S6

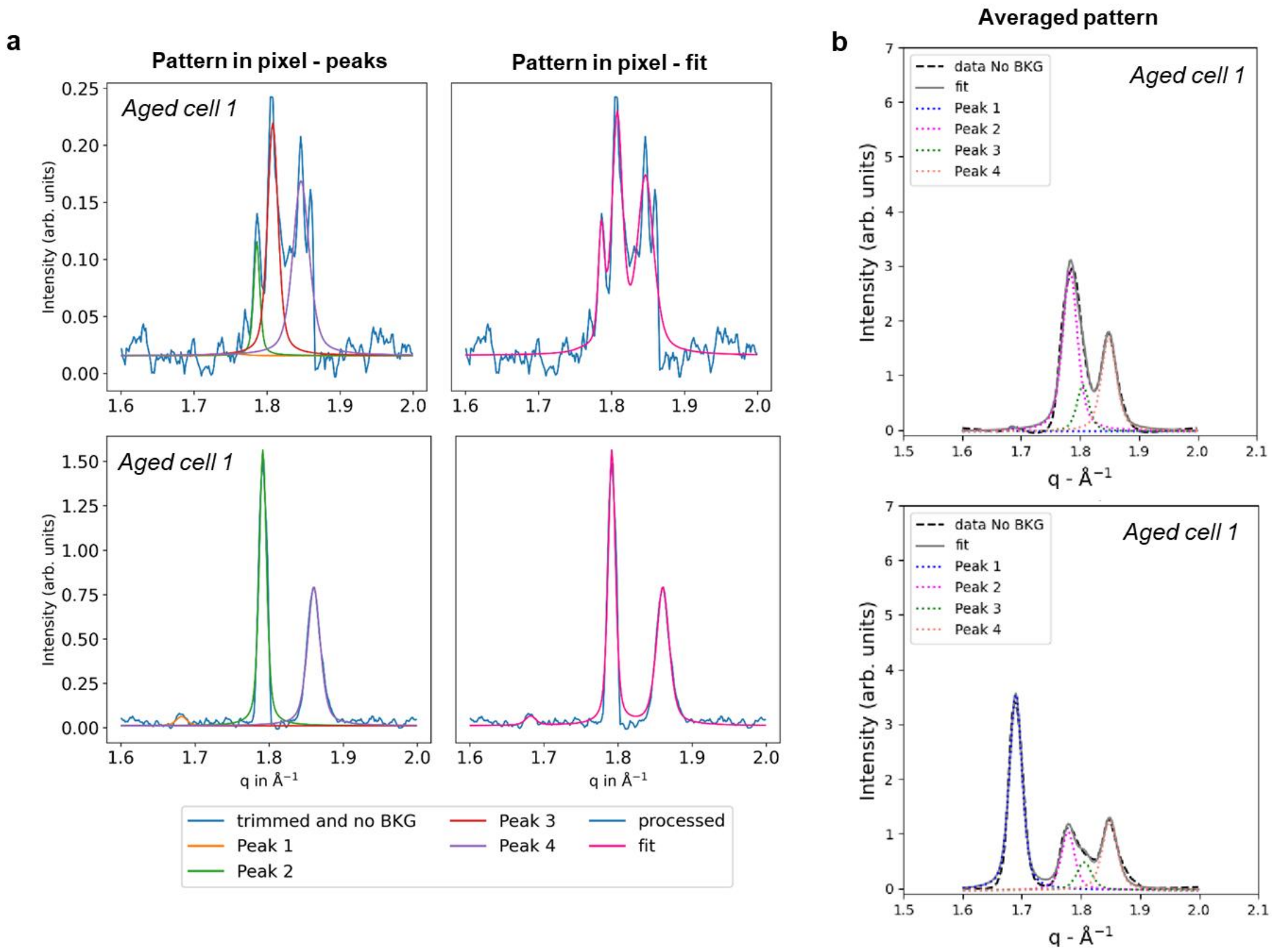

**Figure S8** (a) Example of peak fitting for two selected individual patterns corresponding to distinct times and pixels. Blue lines correspond to trimmed; background subtracted, and processed patterns. Peak 1 (orange) corresponds to stage 1, peak 2 (green) to stage 2/2L, peak 3 (red) to stage 3, and peak 4 (purple) to unlithiated graphite (Gr). The overall fit is represented with a pink line. (b) Example of peak fitting applied to spatially averaged XRD patterns of aged graphite during cycling at a C/5 rate. Peak 1 (blue) corresponds to stage 1, peak 2 (pink) to stage 2/2L, peak 3 (red) to stage 3, and peak 4 (orange) to unlithiated graphite (Gr). We note that the noise is much reduced with respect to individual patterns as in A, due to spatial averaging.

Various peak parameters were extracted from the fitting, including position, height, and full width at half maximum (FWHM). The peak area was calculated using the Pseudo-Voigt model as a linear combination of Gaussian and Lorentzian peak shapes as described in Figure S3. The phase percentage distribution is calculated as the ratio of the area under each fitted peak to the total area of all fitted peaks.

**Representativeness of the operando results.**

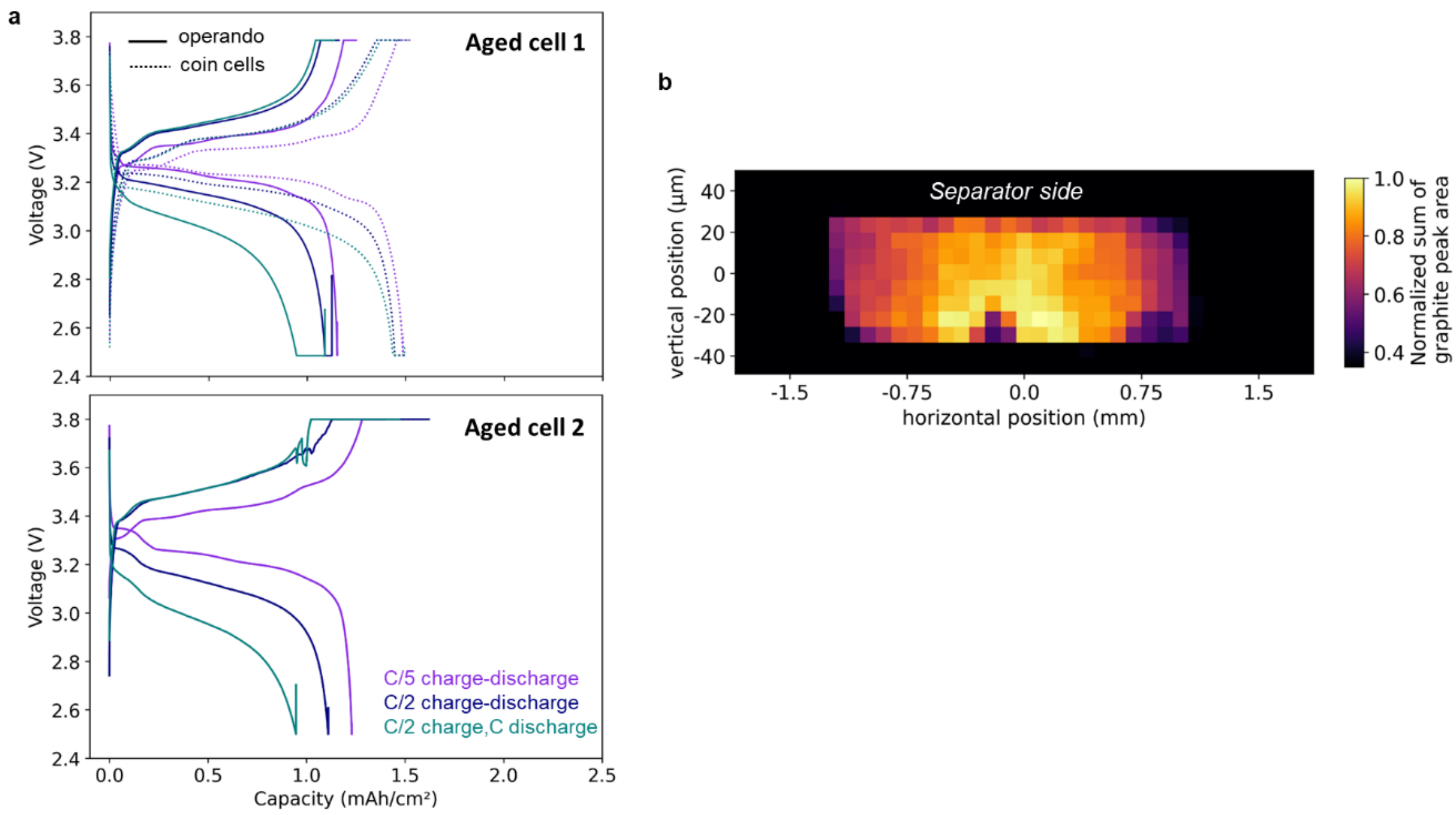

**Figure S9** (a) Galvanostatic cycling curves obtained during operando measurements of the aged electrode cell 2 in the Swagelok cell (bottom panel) compared to cell 1 cycled in Swagelok and coin cells. (b) Masked graphite localization map derived from the sum of X-ray diffraction peak area within the 1.6–1.9 Å region of interest.

A second sample of the aged graphite electrode, punched from the same electrode, was mapped using micro X-ray diffraction at a different beamline of the ESRF Synchrotron (ID31). Similar to the main study, maps consisting of 37 × 13 pixels were acquired during operando cycling at C/5, C/2, and C/2 charge–C discharge cycles. The depth resolution (along z-axis) was 8 µm, while the pixel length along the y-axis was 100 µm.

Both operando-aged cells exhibit very similar electrochemical behavior. The discharge capacities at various C-rates for the second aged cell were measured as 1.28, 1.1, and 0.95 mAh, corresponding to C/5, C/2, and C cycles, respectively. This consistent behavior confirms that the operando XRD analysis results can be reliably compared between the two cells.

**Supporting Section 4: Phase heterogeneities and inactivity quantification**

**Representativeness in the pristine formed electrode**. Phase heterogeneities and the amount of inactive graphite/lithiated graphite phases are quantified by µXRD. First, it is important to evaluate the deviations obtained in several locations of the pristine material due to intrinsic materials' variations to be able to ensure that the effects measured on aged pieces of electrodes are really representative, beyond small changes due to local changes in graphite material microstructural features or electrode composition.

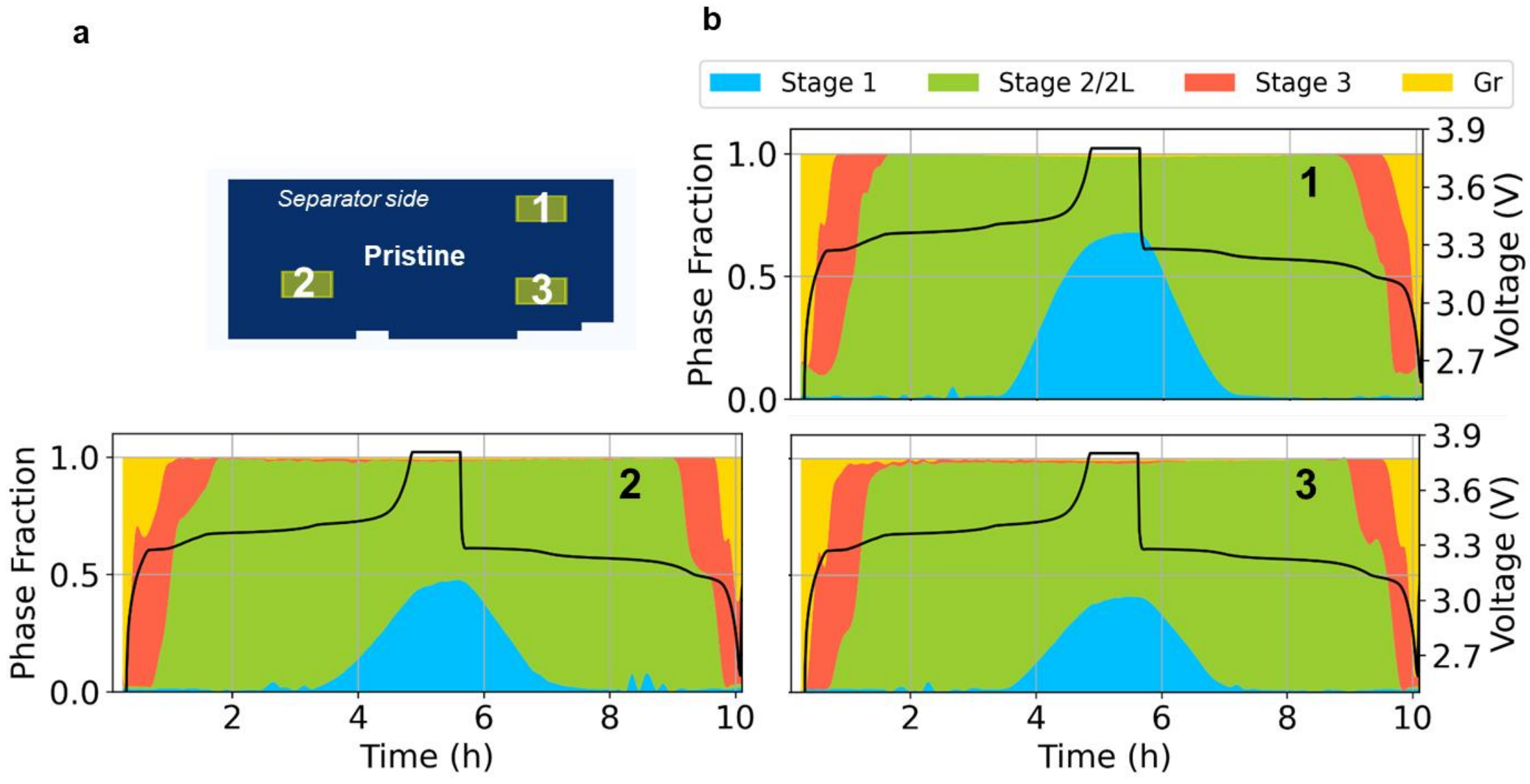

**Figure S10** (a) Schematic representation of the analyzed area on the pristine formed graphite electrode (b) Phase fraction evolution during C/5 charge and discharge in the selected zone of pristine formed graphite. Each color represents a graphite-staging phase identified through X-ray diffraction fitting. Non-lithiated graphite is shown in yellow, stage 3 in coral, stage 2/2L in green, and stage 1 in blue. Three zones are represented (different from the main text results in which the top left is shown): (1) top right near the separator, (2) bottom left close to the current collector, and (3) bottom right close to the current collector.

Figure S8 illustrates the phase fractions during charge and discharge at three distinct locations within the pristine formed electrodes, indicating that different areas behave similarly. However, a notable difference is observed in the stage 1 phase fraction. In the top right region (zone 1), at the end of a C/5 charge, 37% of stage 2/2L and 62% of stage 1 are formed. In contrast, these values decrease to 43/55% and 56/40% for zones 2 and 3, which are closer to the current collector. This corresponds to a global graphite composition of $Li_{0.8}C_6$, $Li_{0.72}C_6$, and $Li_{0.68}C_6$ for zones 1, 2, and 3, respectively. Even though they are close, this suggests heterogeneity in the graphite lithiation between the separator and the current collector side in the pristine formed electrode.

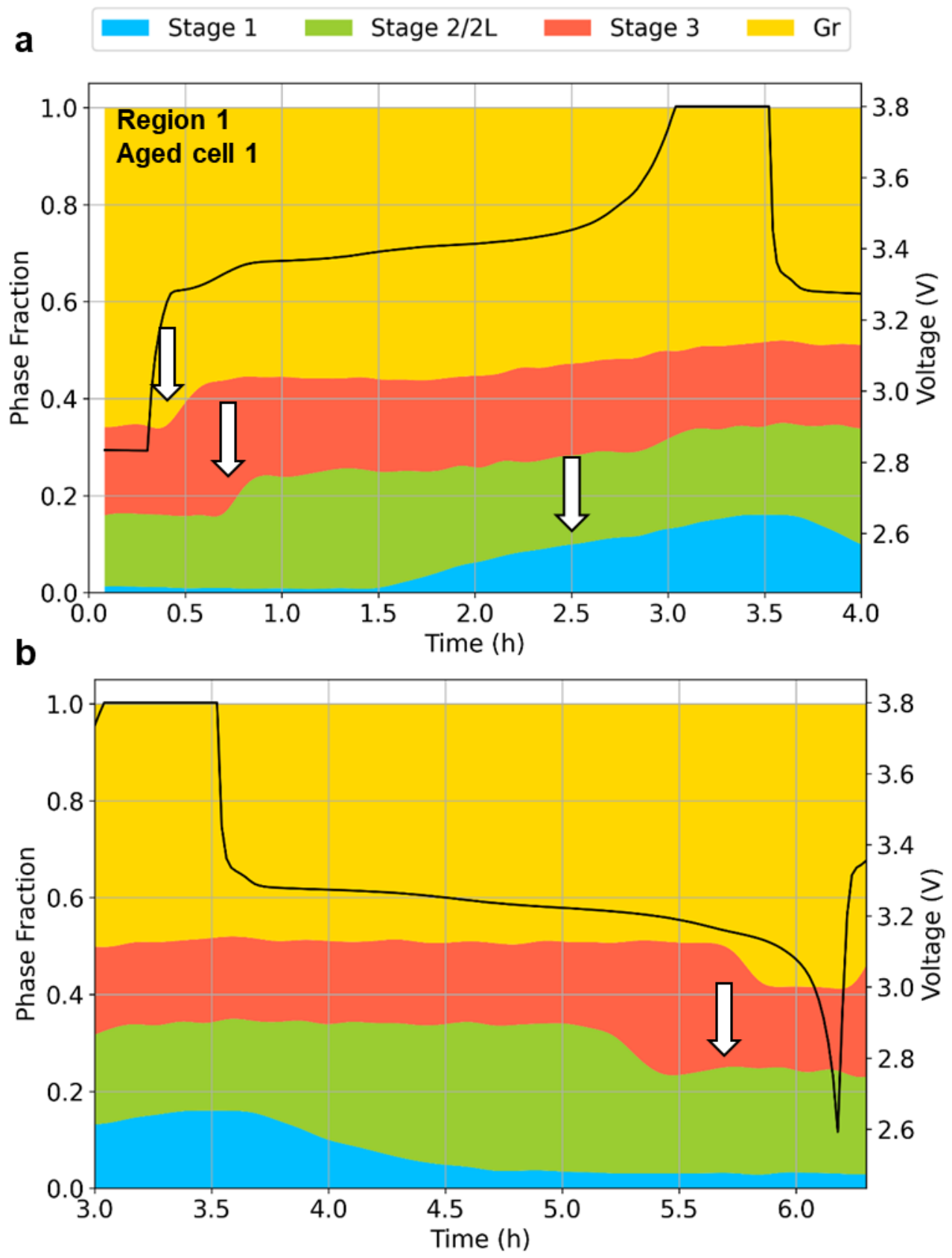

**Figure S11** Phase fraction evolution during C/5 charge (a) and discharge (b) in the selected zone 1 of aged graphite (Cell 1). White arrows indicate the highlighted phenomena during (de)lithiation.

In Figure S9, during charge, only 10% of the graphite transitions into stage 3, stage 2/2L, and stage 1. A continuous and delayed delithiation of particles persists until the end of the charge. Similarly, a delayed lithiation is observed during discharge. Toward the end of the discharge process, lithiation progresses gradually to the stage 2/2L phases.

**XRD patterns of aged electrodes showing the presence of inactive phases.**

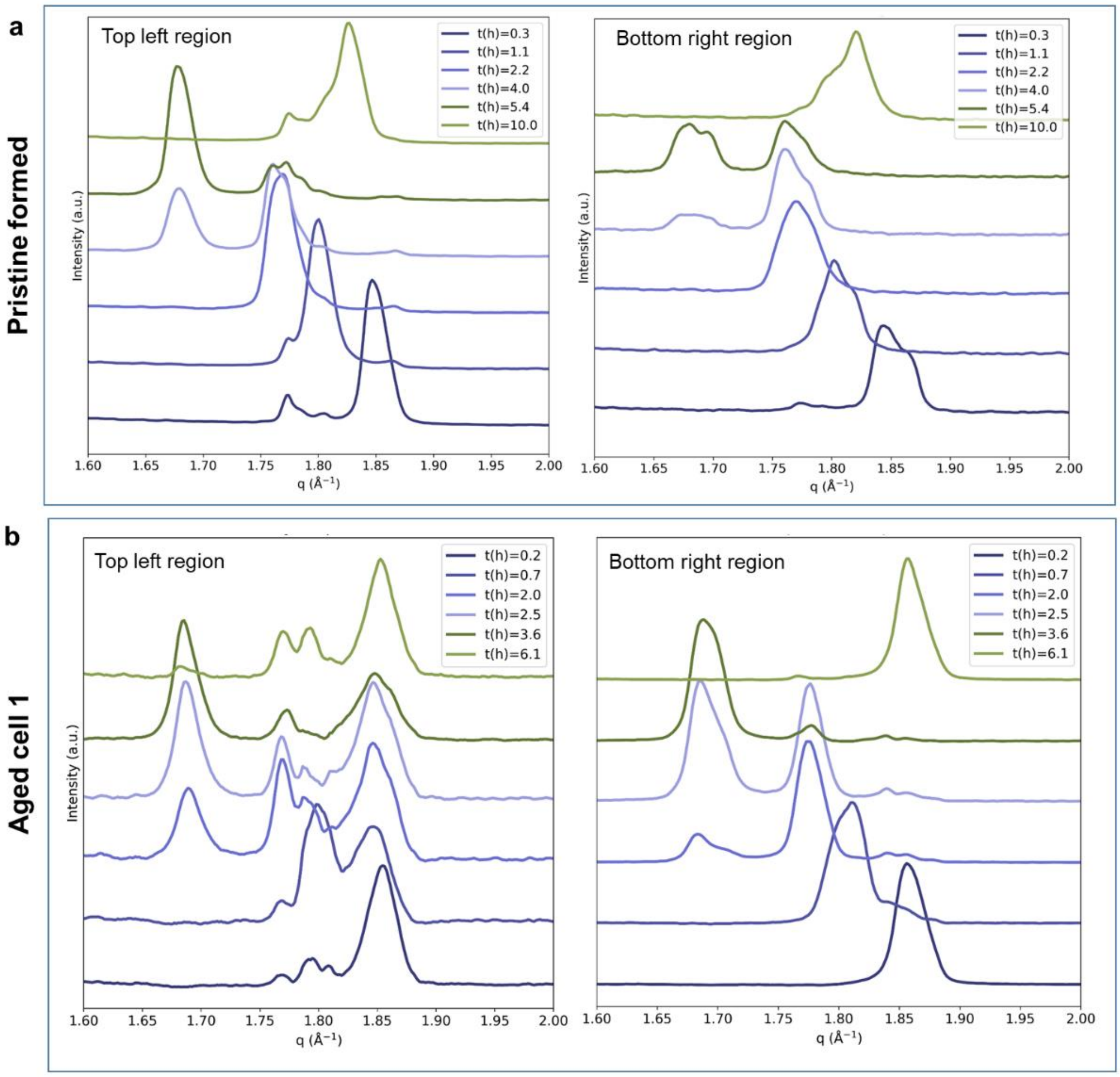

**Figure S12** X-ray diffraction patterns corresponding to the analyzed 3 x 3 pixel-sized regions selected from the pristine formed (a) and aged graphite cell 1 (b) for top left (left) and bottom right (right) regions of the electrodes. The end of charge at C/5 corresponds to t(h) = 5.4 for pristine formed while 3.6 for aged cell 1. t(h)=10.0 and 6.1 represent the end of discharge at C/5 respectively for pristine formed and aged cell 1.

The analysis of X-ray patterns in the selected regions of the pristine formed electrode reveals the expected progression from the unlithiated graphite phase to stage 1. Notably, no stuck phases are observed during this transition. For instance, at t(h)=4, all graphite is partially or fully lithiated. By the end of the charge cycle, the electrode exhibits a mixture of stage 3 and the unlithiated stage.

In contrast, for the aged cell 1, a significant discrepancy is observed between the top-left and bottom-right regions. A substantial amount of unlithiated and partially lithiated graphite phases persists throughout the operando evolution. However, the behavior of the bottom-right region more closely resembles that of the pristine formed electrode. These data highlight the greater level of heterogeneity in the aged electrode.

**Quantifying the total amounts of inactive phases and resolving them spatially in the electrode.**

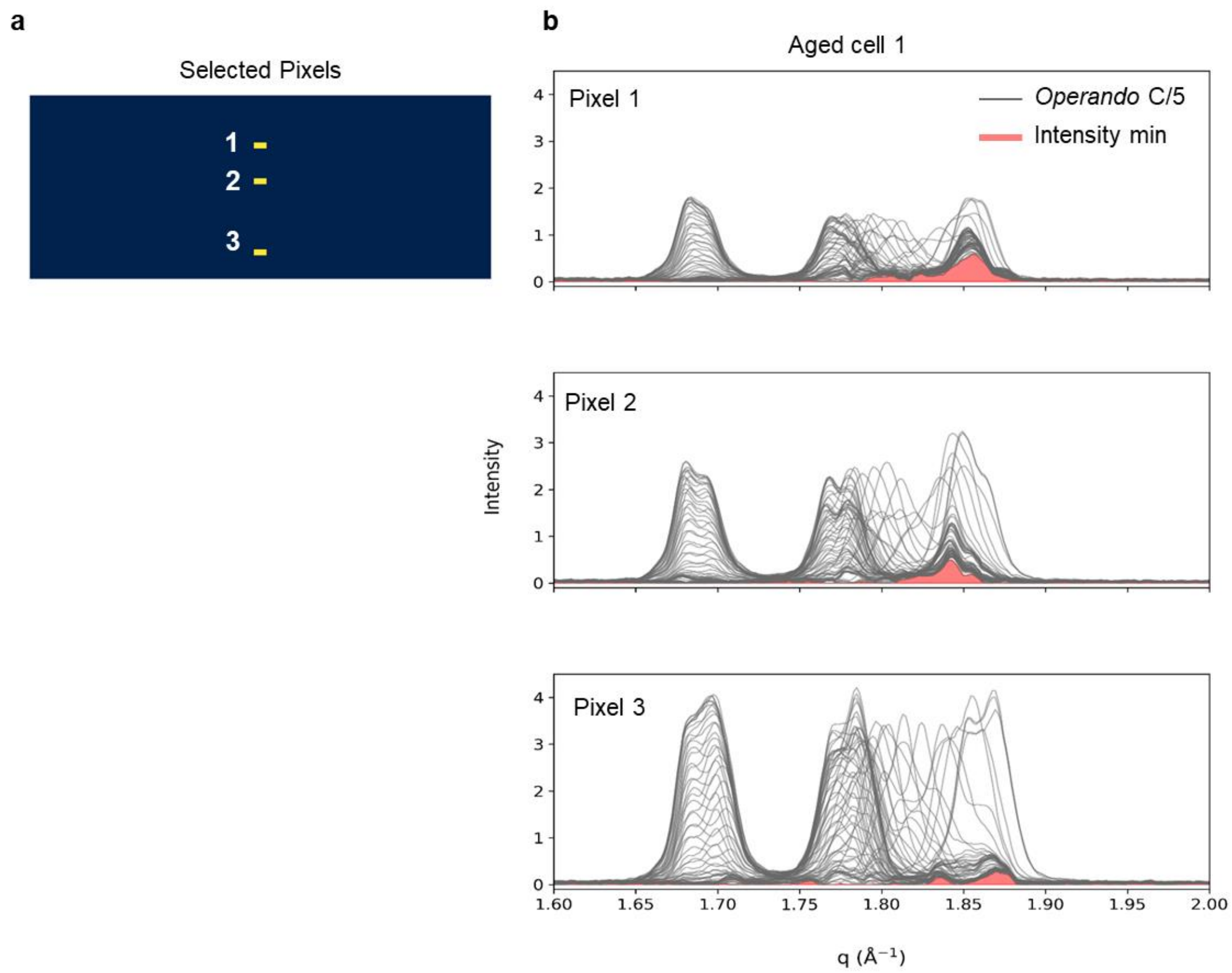

**Figure S13** The calculation of electrochemical inactivity in individual pixels of aged cell 1. (a) Selected exemplary pixels, marked as 1, 2, and 3, were chosen along the same z-line in the sample. (b) The evolution of the X-ray diffractograms (shown in grey) in the region of interest during the C/5 operando sequence. The calculated minimum intensity diffractogram is highlighted in pink.

To obtain the minimum diffractogram,

the minimum filter from the pandas[4] Python library was applied over the selected time sequence. The resulting data, composed of minimum intensity and *q*-values, was subsequently fitted using the PRISMA[3] tool, as detailed in Figure S6. The minimum intensity signal is the part of each diffractogram that does not evolve with time. It might contain pure graphite or lithiated graphite contributions. Hence, this approach enables the quantification of inactive phase fractions for each pixel within the selected time sequence.

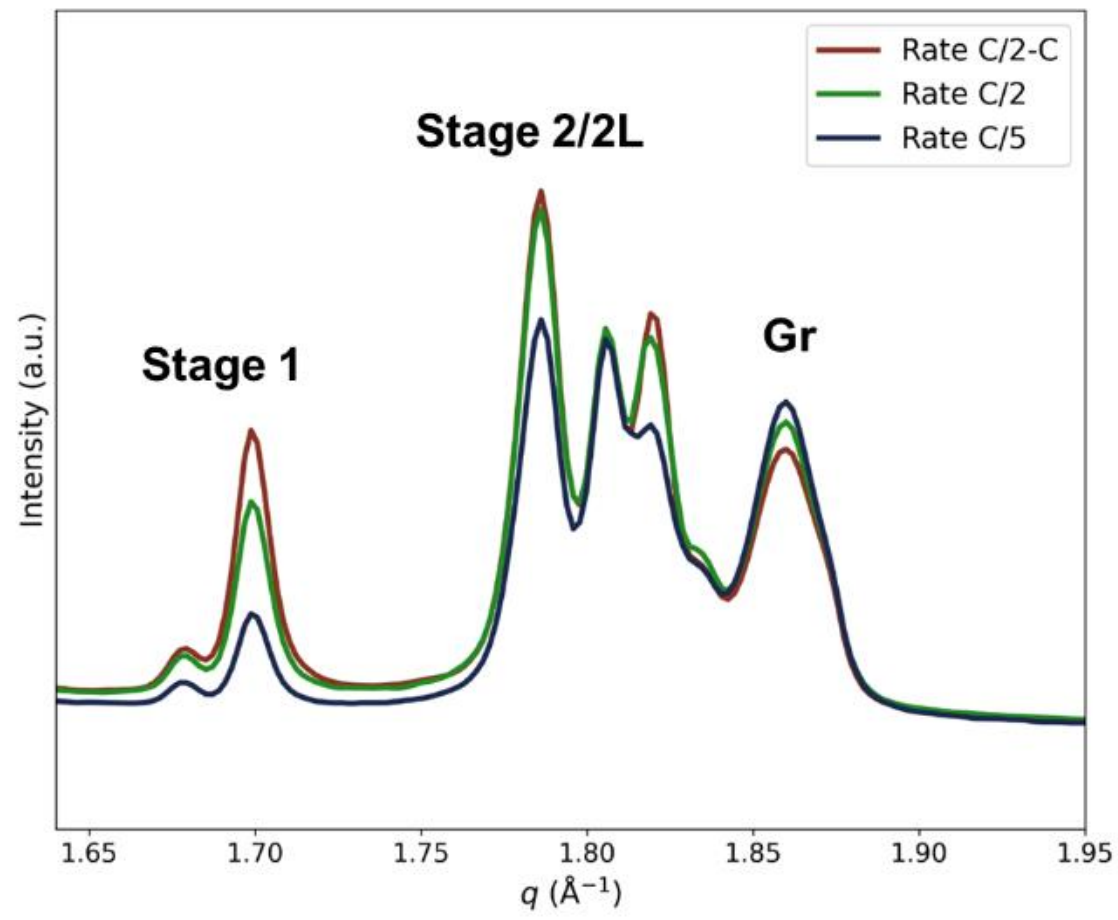

**Figure S14** X-ray diffraction patterns corresponding to isolated inactive graphite phases for different cycling conditions: C/5 cycle in blue, C/2 cycle in green, and C/2 charge - C discharge cycle in red for the aged cell 2.

In Figure S12, the small-intensity diffraction peak at 1.68 $Å^{-1}$ is visible, corresponding to the (100) $LiC_6$ reflection, due to the higher resolution obtained at the ID31 beamline. The most intense peak in this region arises from the (001) $LiC_6$ reflection at 1.69 $Å^{-1}$.

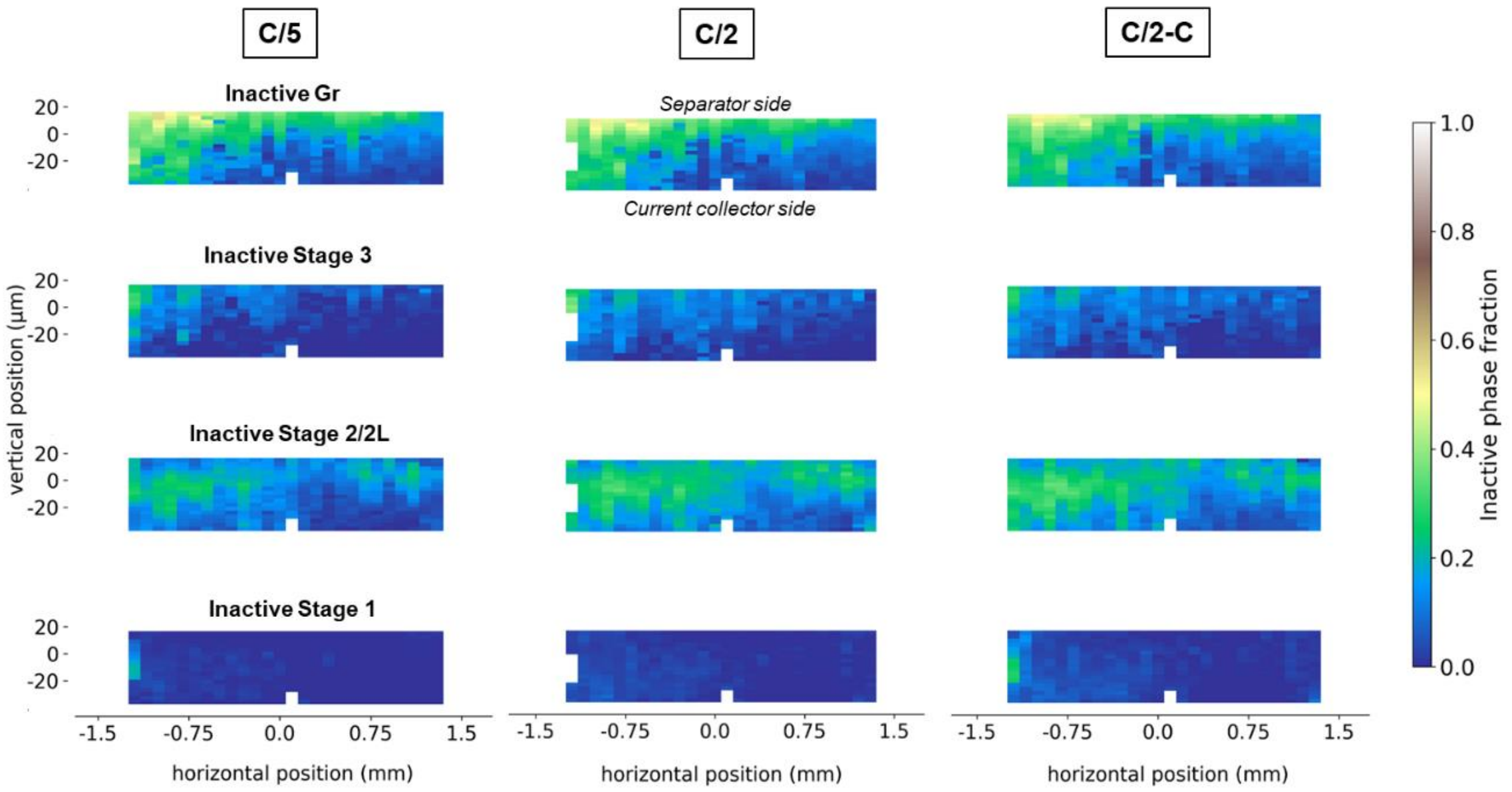

**Figure S15** Inactive phase maps of the aged graphite electrode in cell 1, showing the spatial distribution of averaged inactivity at different C-rates. From top to bottom: Inactive graphite, Inactive stage 3, Inactive stage 2/2L and Inactive stage 1. The color bar represents the inactive phase fraction averaged along the x-axis, and here, 1 corresponds to the total inactivity detected at a given pixel, and 0 indicates that the selected phase is cycling as expected.

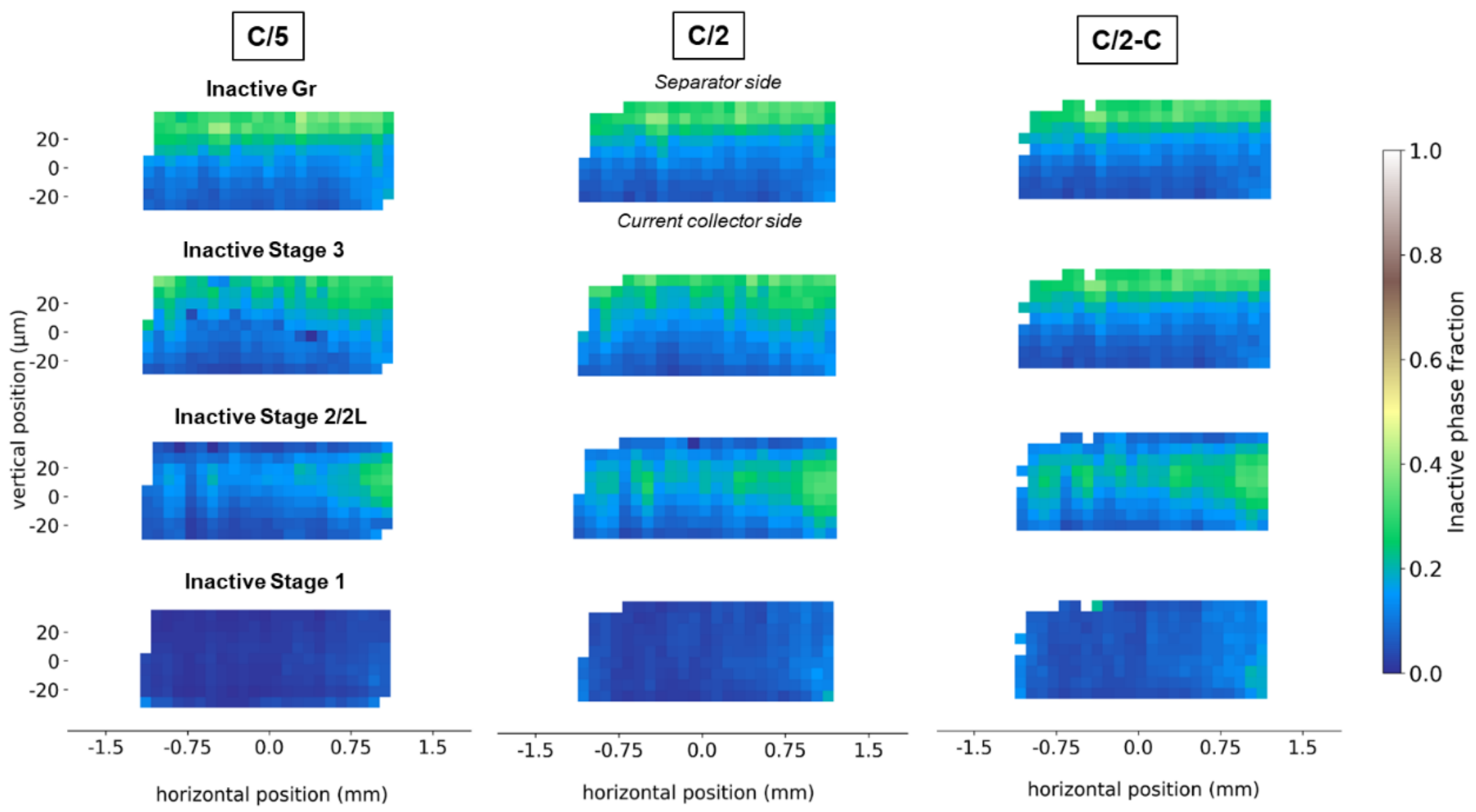

**Figure S16** Inactive phase maps of the second aged graphite electrode in cell 2, showing the spatial distribution of averaged inactivity at different C-rates. From top to bottom: Inactive graphite, Inactive stage 3, Inactive stage 2/2L and Inactive stage 1. The color bar represents the inactive phase fraction averaged along the x-axis, and here, 1 corresponds to the total inactivity detected at a given pixel, and 0 indicates that the selected phase is cycling as expected.

**Supporting Section 5: Lithium concentration quantification and heterogeneity factor**

Using the *q*-values obtained from the center of mass of each peak, the Li concentration was calculated based on previously published refinement results for different lithiated graphite phases. Our group described this approach in detail in a previous publication (Tardif et al.[5]).

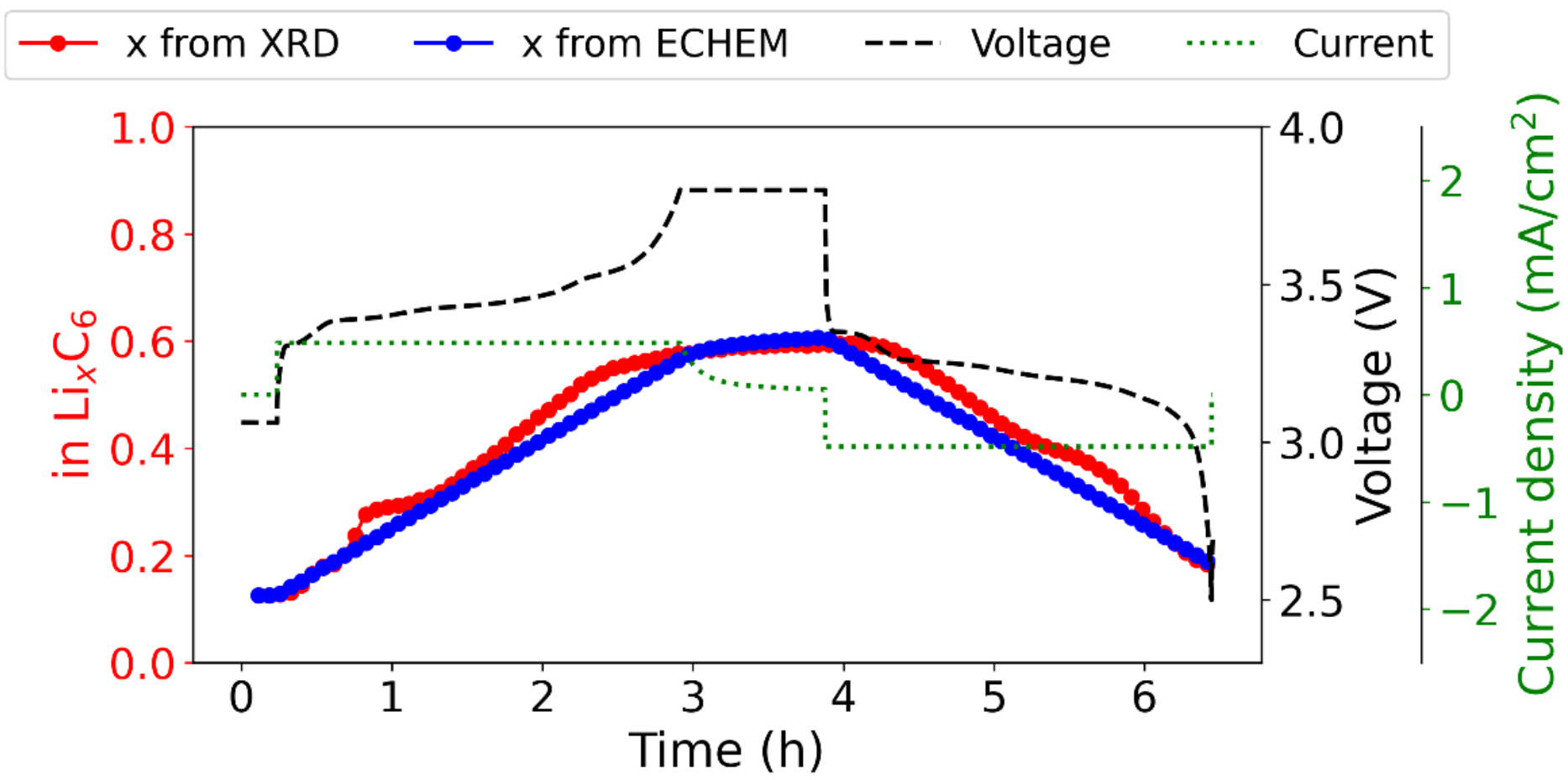

**Figure S17** Evolution of overall lithium content as determined by XRD fitting (red) compared to the lithium amount calculated from delivered capacity during the first charge and discharge cycles at a C/5 rate for the aged cell 2.

The calculated maximum x values for the second aged graphite (cell 2) $Li_xC_6$ are 0.61 from electrochemistry and 0.595 from micro XRD, validating the methodology.

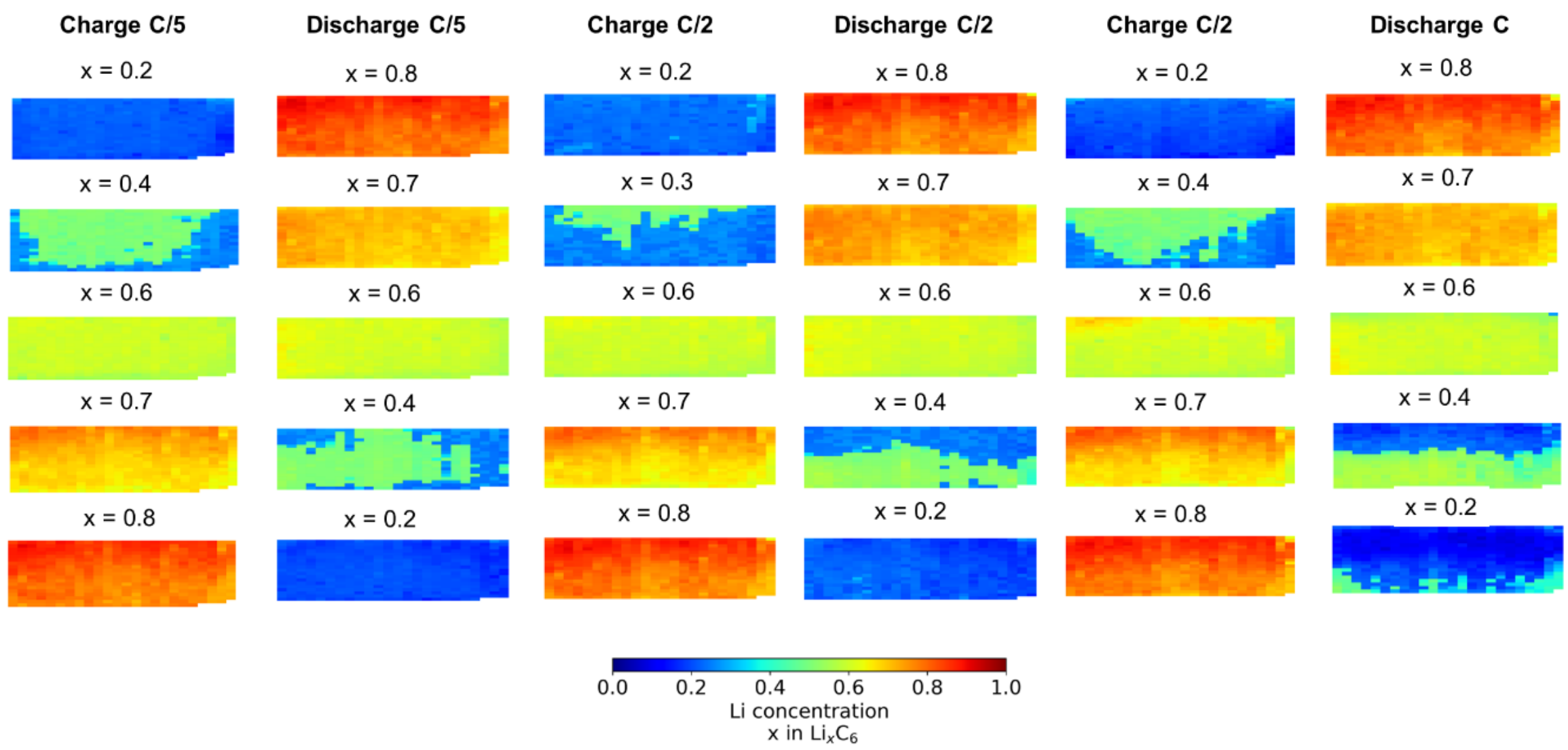

**Figure S18** Lithium concentration maps obtained during operando cycling of the pristine formed electrode. The color map represents the lithium index, denoted as x in $Li_xC_6$. The separator is situated on top of the electrode.

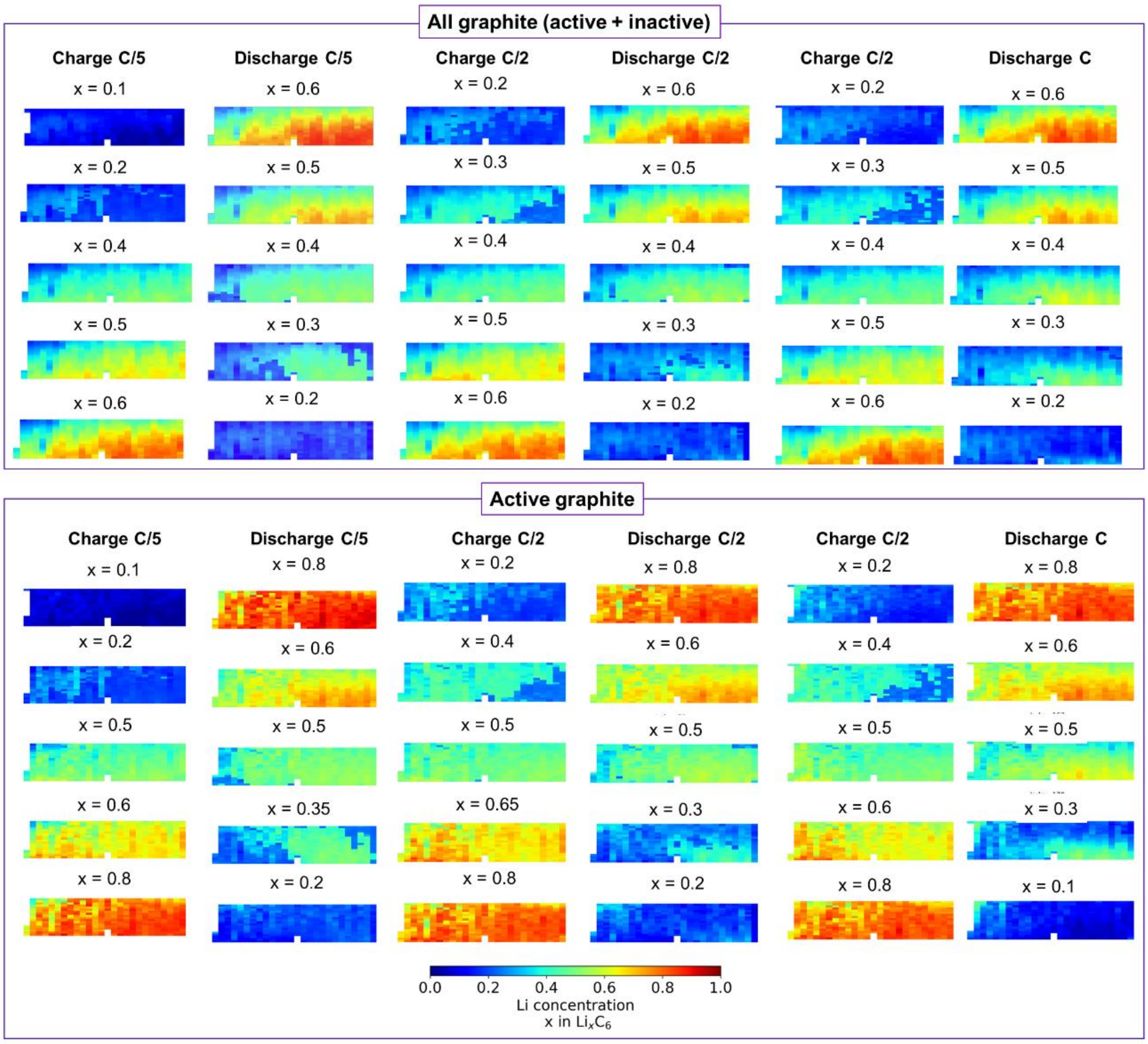

**Figure S19**: Lithium concentration maps obtained during operando cycling in the aged graphite (cell 1) electrode: considering all active and active graphite (top panel) or only the active part of the electrode (bottom panel shown through lithium concentration maps obtained during operando cycling. The color map represents the lithium index, denoted as x in $Li_xC_6$.

## Definition of the heterogeneity factor

The heterogeneity factor is used to quantify local deviations in lithium concentration across the electrode and compare changes in (de)lithiation dynamics. In this study, we defined the time-dependent heterogeneity factor as the spatial integration of local Li concentration deviation at a chosen period of time:

$$Heterogeneity\ (t) = \frac{1}{A}\iint abs(x_{Li}(y,z)(t) - \langle x_{Li}\rangle(t))dydz \qquad (3)$$

where A is the electrode area, $x_{Li}(y,z)$ is the lithium concentration at position (y,z) in the 2D maps, $\langle x_{Li}\rangle$ is the mean electrode lithium concentration (e.g. the value obtained by averaging all pixels in y and z) and (t) is the time corresponding to the lithiation state of the graphite (i.e. state of charge).

**Supplementary References**